\documentclass[aps,prb,twocolumn,groupedaddress,floatfix]{revtex4-2}

\usepackage{graphicx}
\usepackage{dcolumn}
\usepackage{bm}
\usepackage{amsmath}
\usepackage{xcolor}
\usepackage{braket}

\usepackage{bm}
\usepackage[utf8]{inputenc}
\usepackage[T1]{fontenc}
\usepackage{hyperref}
\hypersetup{
breaklinks=true,
    colorlinks=true,
    linkcolor=blue,
    citecolor=blue,
    filecolor=magenta,
    urlcolor=cyan,
}

\usepackage{amsfonts}
\usepackage{blkarray}
\usepackage{amssymb}
\usepackage{multirow}
\usepackage{mathrsfs}

\begin{document}

\title{Molecular formations and spectra due to electron correlations in three-electron 
hybrid double-well qubits}
\author{Constantine Yannouleas}
\email{Constantine.Yannouleas@physics.gatech.edu}
\author{Uzi Landman}
\email{Uzi.Landman@physics.gatech.edu}

\affiliation{School of Physics, Georgia Institute of Technology,
             Atlanta, Georgia 30332-0430}

\date{22 December 2021}

\begin{abstract}
We show that systematic full configuration-interaction (FCI) calculations enable prediction of the energy 
spectra and the intrinsic spatial and spin structures of the many-body wave functions as a function of the
detuning parameter for the case of three-electron hybrid qubits based on GaAs asymmetric double quantum 
dots. Specifically, in comparison with the case of weak interactions and treating the entire three-electron
double-dot hybrid qubit as an integral unit, it is shown that the predicted spectroscopic patterns, 
originating from strong electron correlations, manifest the formation of Wigner molecules (WMs).
Signatures of WM formation include: (1) a strong suppression of the energy gaps relative to the 
non-interacting-electrons modeling, and (2) the appearance of a pair of avoided crossings arising between 
states associated with two-electron occupancies in the left and right wells. The Wigner molecule is a 
physical entity associated with electron localization within each well and it cannot be captured by the 
previously employed independent-particle or two-site-Hubbard theoretical modeling of the hybrid qubits. The 
emergence of strong WMs is investigated in depth through the concerted use of FCI-adapted diagnostic tools 
like charge and spin densities, as well as conditional probability distributions. Furthermore, the energy 
spectrum as a function of the strength of the Coulomb repulsion (at constant detuning) is calculated in 
order to complement the thorough analysis of the factors contributing to WM emergence. We report remarkable
agreement with recent experimental measurements. The present FCI methodology for multi-well quantum dots
can be straightforwardly extended to treat valleytronic two-band Si/SiGe hybrid qubits, where the central 
role of the WMs was confirmed recently. Such valleytronic FCI could be also adapted and employed in 
simulations of Si-based two-qubit logical gates, made of two interacting DQD hybrid qubits confining a 
total of $N \leq 6$ electrons.
\end{abstract}

\maketitle

\section{Introduction}
\label{intr}

Methodical control of the parameters and performance of qubits is a prerequisite for the succesful 
implementation of quantum computing. To this effect during the last decade, major experimental endeavors 
(see, e.g., Refs.\ \cite{kouw07,dzur13,marc10,dzur10,yaco18,marc13,pett21}) 
have been undertaken and substantial 
progress has been reported. In particular, unprecedented progress has been achieved in the techniques for 
controlling and manipulating the spin and charge electronic degrees of freedom of two-dimensional (2D) 
semiconductor hybrid-double-quantum-dot (HDQD) qubits 
\cite{copp12,copp14.2,copp15,cao16,copp17,cao17,erca21,kim21}, comprising three-electron (3e) 
\cite{copp12,copp14.2,copp15,copp17,cao17,kim21} and five-electron (5e) \cite{cao16,erca21} varieties.

Nonetheless, several experimental scrutinies in the last two years on state-of-the-art semiconductor 
double-dot Si/SiGe \cite{erca21} and GaAs \cite{kim21} HDQD qubit devices have provided incontrovertible
evidence (see also Refs.\ \cite{erca21.2,urie21}) that a key factor influencing the qubit spectra and
performance had been overlooked in the context of earlier qubit investigations. This factor is the 
manifestation of strong electron-electron (e-e) correlations leading to formation of Wigner molecules (WMs) 
\cite{yann99,grab99,yann00,yann00.2,boni01,yann02,mikh02,szaf03,cift05,yann06,yann06.2,yann07,yann08,ront08},
which rearranges sharply the electronic spectra of the qubit device with respect to those associated with 
non-interacting electrons. As their name suggests, the WMs are a fully quantal extension of the concept
of a bulk Wigner crystal \cite{wign34} to the realm of finite systems.

From a theoretical perspective, the formation of Wigner molecules cannot be described in the framework of
independent-particle (single-particle) modeling \cite{copp12,copp14,vinc15}, which was advanced for 
2D quantum dots (QDs) from the very early stages of the field \cite{marc96}; nor the more involved 
{\it two-site} Hubbard models \cite{copp12,staf94,loss99,dass11,dass11.2,ferr14,burk17} are adequate in this
respect \cite{note3}. As reaffirmed recently for the case of two particles (two electrons 
\cite{kim21,erca21,erca21.2,urie21} or two holes \cite{urie21}) in a single QD, explored in recent 
experimental investigations \cite{kim21,erca21}, formation of WMs, which has been extensively demonstrated 
in earlier investigations \cite{yann99,yann00,yann00.2,yann02,yann06,yann06.2,yann07,yann07.2,yann08,yann09},
requires the employment of more comprehensive, {\it ab-initio\/}-in-nature, theoretical approaches, 
including the symmetry-breaking/symmetry-restoration \cite{yann99,yann00,yann02,yann07} methods and
the full configuration-interaction (FCI) method (referred to also as exact diagonalization,
\cite{yann00.2,yann06,yann06.2,yann07,yann07.2,yann08,yann09,shav98,note1}).

In this paper, we demonstrate the central role that strong e-e correlations and WM formation play in 
shaping the spectra of semiconductor qubits by going beyond the aforementioned recent two-particle CI 
calculations in a single dot \cite{erca21,kim21,erca21.2,urie21}, namely, we present FCI calculations for 
the case of a hybrid \cite{erca21,kim21,copp12,copp14,copp14.2,copp15,cao16,copp17,cao17} 
{\it three-electron\/} double-quantum-dot GaAs qubit, treated as an integral unit.
To establish contact with an actual experimental 
investigation, we employ HDQD parameters comparable to those in Ref.\ \cite{kim21}, which measured
the energy spectrum of the qubit as a function of the detuning at zero magnetic field.
As we elaborate below, we find a remarkable agreement between our 3e-CI energy spectra and the essential 
experimental trends that characterize the hybrid qubit. These trends include the large suppression of the 
energy gaps compared to the simple non-interacting-electron problem and the emergence of the 
phenomenological 4x4 effective Hamiltonian, which incorporates a pair of avoided crossings associated with
double electronic occupancies in the left and right well, and which depends linearly on the detuning
between the two wells.  

We further present a detailed analysis of the interplay of the spectral features of the HDQD and the
formation of a WM, by investigating charge and spin-resolved densities, as well as spin-resolved 
conditional probability distributions (CPDs), for the 6 or 8 lowest-in-energy states of the 3e-DQD 
spectra for two different cases, namely as a function of detuning (keeping constant the dielectric 
constant $\kappa$) and as a function of the dielectric constant (keeping the detuning constant). 
In this context, particularly
revealing for the process of WM formation is the contrast of charge densities (see Fig.\ \ref{spccp} 
below) between a weakly interacting case (with $\kappa=1000$) and the strongly interacting case (with 
$\kappa=12.5$) of the GaAs HDQD qubit.

{\it Plan of the paper:}
In Sec.\ \ref{sec:mbh}, we introduce the many-body Hamiltonian of the HDQD device model, comprising 
three electrons in an asymmetric 2D double-well external confinement, which is modeled by a 
two-center-oscillator (TCO) potential joined by a smooth neck. The values of the model's parameters, 
employed in the calculations, are specified in this section; they are chosen to address the experiments 
on GaAs HDQD in Ref.\ \cite{kim21}. Also included (Sec.\ \ref{sec:wigpar}) in this section is a 
restatement of the Wigner parameter $R_W$ \cite{yann99,yann07} pertaining to the propensity for 
Wigner-molecule formation, as well as estimates of the $R_W$ corresponding to the parameters used in our 
calculations.
 
In Sec.\ \ref{sec:specdet}, the CI-calculated energy spectra as a function of the interdot detuning 
parameter at specific values of the dielectric constant (controlling the strength of the Coulomb 
repulsion) are discussed (Sec.\ \ref{sec:big}) for both the case where the inter-electron interaction is 
taken to be small (simulating the independent-electron limit) and for the actual value appropriate for 
GaAs (illustrating the effect of strong WM-formation on the eigenvalue spectra). Subsequently (Secs.\ 
\ref{sec:char1} and \ref{sec:spnstr}), we analyze the charge densities and spin structures corresponding 
to selected values of the detuning parameter away from the avoided crossings, using both spin-resolved 
charge densities and spin-resolved CPDs. Special attention is devoted in Sec.\ \ref{sec:avcr} to the 
evolution of the spectral characteristics and the mixing of WM wave functions (Sec.\ \ref{sec:mix}) when 
the detuning values fall in the neighborhood of the avoided crossings. In Sec.\ \ref{sec:gap}, we focus 
on the energy gap between the ground and 1st excited state, and report remarkable agreement with the 
measurements of Ref.\ \cite{kim21}. We end Sec.\ \ref{sec:specdet} with an analysis (Sec. 
\ref{sec:effham}) of the CI-calculated spectra for GaAs in the context of a phenomenological matrix 
Hamiltonian used as a diagnostic tool in previous works on HDQD qubits.  

The effects of inter-electron interaction on the Wigner-molecule formation and associated spectral 
characteristics, are further highlighted and accentuated in Sec.\ \ref{spectdie} through the analysis of 
the CI-calculated spectra and charge densities as a function of the dielectric constant of the material 
at a constant detuning value. We summarize our results in Sec.\ \ref{sec:sum}.

Appendix \ref{tco} describes the numerical approach to determine the eigenenergies and eigenstates of 
the one-body TCO Hamiltonian introduced below in Eq.\ (\ref{hsp}). Because the 
community of quantum-information and quantum-computer scientists have only recently become alerted 
\cite{erca21,erca21.2,kim21,urie21} to the potentialities of the CI many-body method and the 
significance of the concept of the Wigner molecule, for completeness and pedagogical reasons, we 
include two additional Appendices as follows:
Appendix \ref{cime} describes the CI methodology, whereas Appendix \ref{spdcpd} describes the diagnostic 
tools (beyond-mean-field single-particle densities and CPDs, and their spin-resolved varieties) needed 
to analyze the CI many-body wave functions and extract the information regarding WM formation and
the intrinsic spin structure of the associated CI wave function.       

\section{Many-body Hamiltonian and parameters of the double-well device}
\label{sec:mbh}

Following the recent advances \cite{kim21,erca21,cao16,cao17} in the fabrication of hybrid qubits, 
we investigate in this paper the many-body spectra and wave functions of three electrons in an 
asymmetric two-dimensional double-well external confinement, inplemented by a 
two-center-oscillator (TCO) potential as described below. 

\begin{figure*}
\centering\includegraphics[width=17cm]{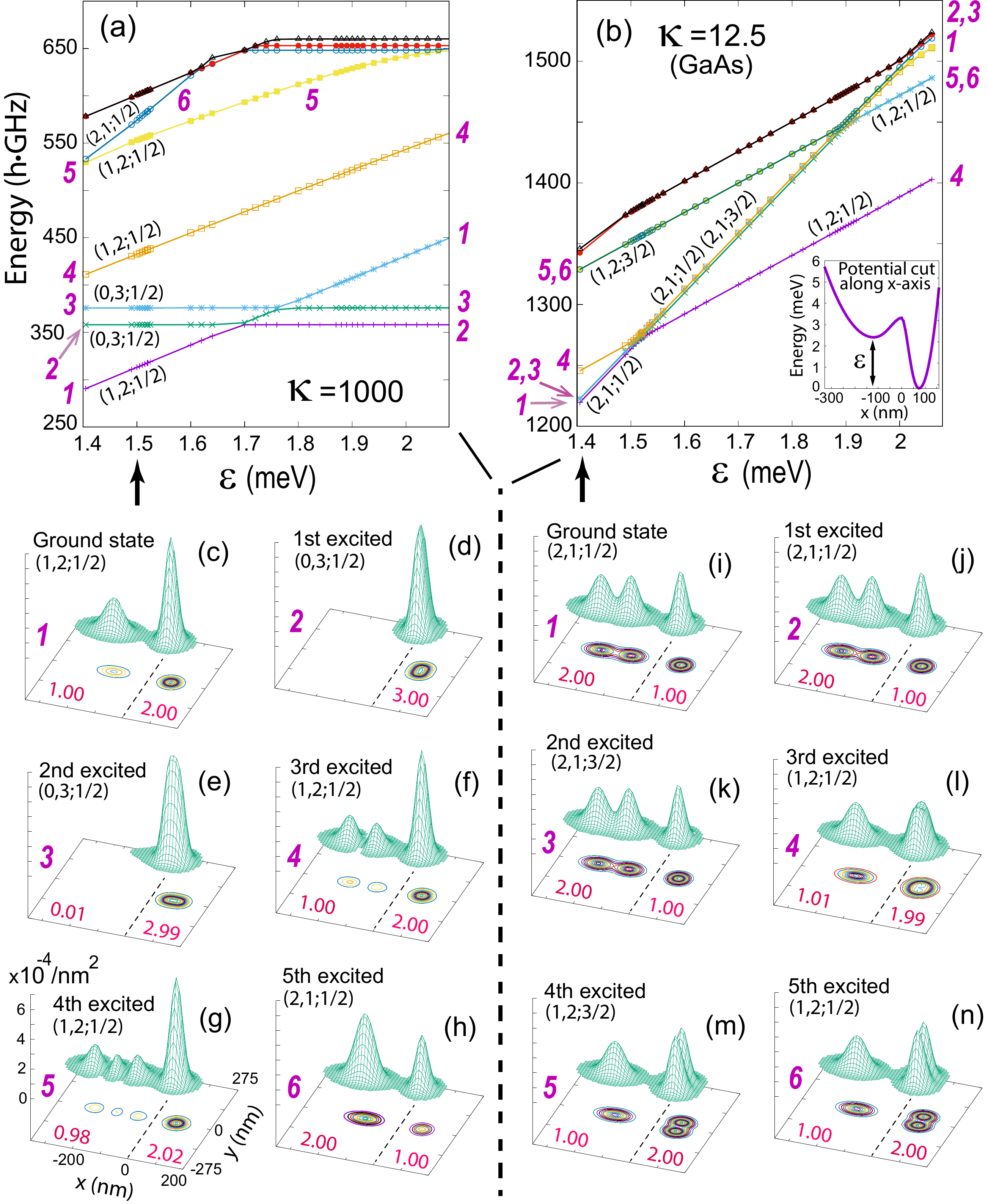}
\caption{
Spectra for the same three-electron double dot for two different values of the dielectric
constant.
(a) $\kappa=1000$ (weak Coulombic repulsion). (b) $\kappa=12.5$ (actual value for GaAs,
stronger Coulombic repulsion). (c-h) Charge densities for the ground and first five excited states
for $\kappa=1000$. (i-n) Charge densities for the ground and first five excited states for
$\kappa=12.50$. The arrows indicate the value of the detuning at which the charge
densities were calculated, i.e., $\varepsilon=1.50$ meV for (c-h) and $\varepsilon=1.405$ for 
(i-n). The notation $(n_L, n_R; S)$ denotes the left electron occupation, the right electron
occupation, and the total spin, respectively. For all densities, the scales of all three axes are
as in (g). CI left and right occupations are highlighted in red.
In all figures and in the text, the CI energies have been referenced to 
$3 \hbar \sqrt{(\omega_{x1}^2 + \omega_{y}^2)/2 }$.
}
\label{spccp}
\end{figure*}

We consider a many-body Hamiltonian for $N$ electrons of the form
\begin{equation}
{\cal H}_{\rm MB} ({\bf r}_i,{\bf r}_j) =\sum_{i=1}^{N} H_{\rm TCO}(i) +
\sum_{i=1}^{N} \sum_{j>i}^{N} V({\bf r}_i,{\bf r}_j),
\label{mbhd}
\end{equation}
where ${\bf r}_i$,${\bf r}_j$ denote the vector positions of the $i$ and $j$ electron. This Hamiltonian 
is the sum of a single-particle part $H_{\rm TCO}(i)$, which implements the double-well confinement,
and the two-particle interaction $V({\bf r}_i,{\bf r}_j)$. A notable property of $H_{\rm TCO}$ is the
fact that it allows for the formation of a smooth interwell barrier between the individual wells; see
the inset of Fig.\ \ref{spccp}(b) for an illustration. 

Naturally, for the case of electrons, the two-body interaction is given by the Coulomb repulsion
\begin{equation}
V({\bf r}_i,{\bf r}_j)=\frac{e^2}{\kappa |{\bf r}_i-{\bf r}_j|},
\label{tbie}
\end{equation}
where $\kappa$ is the dielectric constant of the semiconductor material. 

In the two-dimensional TCO employed by us, the single-particle levels associated with the 
confining potential are determined by the single-particle hamiltonian \cite{yann99,yann02,yann09}
\begin{equation}
H_{\rm TCO}=\frac{{\bf p}^2}{2 m^*} + \frac{1}{2} m^* \omega^2_y y^2
    + \frac{1}{2} m^* \omega^2_{x k} x^{\prime 2}_k + V_{\rm neck}(x) +h_k,
\label{hsp}
\end{equation}
where $x_k^\prime=x-x_k$ with $k=1$ for $x<0$ (left) and $k=2$ for $x>0$ (right), 
and the $h_k$'s control the relative depth of the two wells, with the detuning
defined as $\varepsilon=h_1-h_2$. $y$ denotes the coordinate perpendicular to the
interdot axis ($x$). The most general shapes described by $H$ are
two semiellipses connected by a smooth neck [$V_{\rm neck}(x)$]. $x_1 <
0$ and $x_2 > 0$ are the centers of these semiellipses, $d=x_2-x_1$
is the interdot distance, and $m^*$ is the effective electron mass.

For the smooth neck, we use 
\begin{align}
V_{\rm neck}(x) = \frac{1}{2} m^* \omega^2_{x k} 
\Big[ {\cal C}_k x^{\prime 3}_k + {\cal D}_k x^{\prime 4}_k \Big] \theta(|x|-|x_k|),
\label{vneck}
\end{align}
where $\theta(u)=0$ for $u>0$ and $\theta(u)=1$ for $u<0$.
The four constants ${\cal C}_k$ and ${\cal D}_k$ can be expressed via two
parameters, as follows: ${\cal C}_k= (2-4\epsilon_k^b)/x_k$ and
${\cal D}_k=(1-3\epsilon_k^b)/x_k^2$, where the barrier-control parameters
$\epsilon_k^b=(V_{b}-h_k)/V_{0k}$ are related to the height of the targeted 
interdot barrier ($V_{b}$, measured from the zero point of the energy scale), 
and $V_{0k}=m \omega_{x k}^2 x_k^2/2$. We note that measured from the bottom of 
the left ($k=1$) or right ($k=2$) well the interdot barrier is $V_{b}-h_k$.

How we solve for the eigenvalues and eigenstates of $H_{\rm TCO}$ is described in Appendix \ref{tco}.

Motivated by the asymmetric double-dot used in the GaAs device described in Ref.\ \cite{kim21},
we choose the parameters entering in the TCO Hamiltonian as follows: The left dot
is elliptic with frequencies corresponding to 
$\hbar\omega_{x1}=0.413567~{\rm meV}=100$ h$\cdot$GHz and $\hbar\omega_{y1} = \hbar \omega_y 
=1.22~{\rm meV}=294.9945$ h$\cdot$GHz (1 h$\cdot$ GHz $=4.13567$ $\mu$eV), whereas the right dot 
is circular with $\hbar\omega_{x2}=\hbar\omega_{y2}= \hbar \omega_y = 1.22~{\rm meV}=
294.9945$ h$\cdot$GHz. The left dot is located at $x_1=-120$ nm, whereas the right dot is 
located at $x_2=75$ nm, and the interdot barrier is set to $V_b=3.3123$ meV $=800.91$ 
h$\cdot$GHz. The effective electron mass and the dielectric constant for GaAs are
$m^*=0.067 m_e$ and $\kappa=12.5$, respectively.

\subsection{The Wigner parameter}
\label{sec:wigpar}

At zero magnetic field and in the case of a single circular harmonic QD, the degree of electron
localization and Wigner-molecule pattern formation can be associated with the socalled Wigner 
parameter \cite{yann99,yann07} 
\begin{align}
R_W =  Q/(\hbar \omega_0),    
\label{rw}
\end{align}
where $Q$ is the Coulomb interaction strength and $\hbar \omega_0$ is the energy quantum of the 
harmonic potential confinement (being proportional to the one-particle kinetic energy); 
$Q=e^2/(\kappa l_0)$, with $l_0=(\hbar/(m^*\omega_0))^{1/2}$ the spatial extension of the lowest
state’s wave function in the harmonic (parabolic) confinement.

Naturally, strong experimental signatures for the formation of Wigner molecules are not expected 
for values $R_W \lesssim 1$. In the double dot under consideration here, there are two different 
energy scales, $\hbar\omega_1=0.413567~{\rm meV}$ (associated with the long $x$ dimension of the 
left QD) and $\hbar\omega_2=1.22~{\rm meV}$ (associated with the right circular QD). As a result, 
for GaAs (with $\kappa=12.5$) one gets two different values for the Wigner parameter, namely 
$R_{W,1}=5.31$ and $R_{W,2}=3.09$. These values suggest that a stronger Wigner molecule should 
form in the left QD compared to the right QD, as indeed was found by the FCI calculation described
below; the essentials of the FCI method are presented in Appendix \ref{cime}. 

We note that the earlier fabricated GaAs quantum dots had harmonic confinements associated with 
frequencies $\hbar \omega_0 \geq 3$ meV ($R_W < 1.97$) \cite{marc96,vinc15}, which correspond to a 
range of smaller QD sizes that did not favor the observation of the WMs at zero magnetic fields, as 
can be concluded from an inspection of the earlier experimental literature \cite{kouw01}. In 
this context, the much larger anisotropic GaAs double dot of Ref.\ \cite{kim21}, as well as
the findings of Ref.\ \cite{erca21}, where strong WM signatures were observed, heralds the 
exploration of till now untapped potentialities in the fabrication and control of quantum dot 
qubits.

\section{CI spectra as a function of detuning}
\label{sec:specdet}

Before engaging in detailed analyses of the CI numerical results, we comment on a particular notation 
that will be essential in facilitating this task. Indeed, we will extensively use the three-part 
notation $(n_L, n_R; S)$ (with $n_L+n_R=N$) to denote the left-well electron occupation, the right-well 
electron occupation, and the total spin, respectively, associated with a 3e CI wave function. In 
this vein, the two-part notation $(n_L, n_R)$ will also be occasionally used

\subsection{The big picture}
\label{sec:big}

Fig.\ \ref{spccp} compares the CI spectra as a function of detuning (within the same window, 
$1.40 \mbox{~meV} \leq \varepsilon \leq 2.1 \mbox{~meV}$) for two different values of the 
dielectric constant, i.e., $\kappa=1000$, which is closer to the non-interacting limit, and 
$\kappa=12.5$, which is the actual value for GaAs QDs. 

This comparison demonstrates a dramatic modification in the spectra. Indeed, the low-energy 
spectrum for $\kappa=1000$ [including the ground state, see Fig.\ \ref{spccp}(a)] is dominated by 
states that have the majority of electrons (two or all three electrons) residing in the deeper 
{\it right} well; such states are denoted as $(1,2;S)$ or $(0,3;S)$. According to the socalled
branching diagram \cite{paunczbook,note2}, for three electrons the allowed total spin values are 
$S=1/2$ (with a multiplicity of two) and $S=3/2$ (with a multiplicity of one). All five 
lowest-in-energy states in Fig.\ \ref{spccp}(a) have a total spin $S=1/2$, with the $S=3/2$ 
states appearing at higher energies (at the top of the plotted energy spectrum). 

On the contrary, in the corresponding low-energy spectrum in the GaAs case [Fig.\ \ref{spccp}(b)],
the $(2,1;S)$ states with two electrons in the {\it left} well are prominent, along the
$(1,2;S)$ states with two electrons in the right well. Furthermore, states with three electrons in
a given well, denoted as $(3,0;S)$ or $(0,3;S)$, are absent. The fact that only the six 
$(2,1;S)$ and $(1,2;S)$ states comprise the lowest-energy spectrum for the GaAs double dot is an
essential feature that is a prerequisite for the implementation of the hybrid qubit which uses
\cite{copp12,copp14,copp14.2,kim21,cao17} the four $(2,1;1/2)$ and $(2,1;1/2)$ states.
As it is discussed below, this feature is the effect of the formation of Wigner molecules as a 
result of the strengthening of the typical Coulomb energies relative to the energy gaps in the 
single-particle spectrum of a confining external potential that represents a rather large-size
and strongly asymmetric double dot (see the discussion on the Wigner parameter $R_W$ in
Sec.\ \ref{sec:wigpar}).      

To assist the reader in the exploration of the features present in the spectra of Fig.\
\ref{spccp}, we have successivley numbered the lowest six states at $\varepsilon=1.4$ meV, starting
from the ground state (\#1) and moving upwards to the first five excited ones. Apart from the
immediate neighbohood of an avoided crossing, in both spectra, these energy curves are straight
lines, and naturally we keep the same numbering for all values of the detuning in the window range
used in Figs.\ \ref{spccp}(a,b).

In Fig.\ \ref{spccp}(a), there are no degeneracies, and this numbering is self-explanatory. 
The spectrum in Fig.\ \ref{spccp}(b) is less transparent, because of quasi-degeneracies between
states \#2 and \#3 and  \#5 and \#6, as well as the small energy gap ($\sim$ 3 h$\cdot$GHz)
between state \#1 and the quasi-degenerate pair (\#2,\#3). We stress that the states \#1 and \#2 
have two electrons in the left well and total spin $S=1/2$, and thus they are denoted as
$(2,1;1/2)$, whereas state \#3 has two electrons in the left well, but a total spin of $S=3/2$
[denoted as $(2,1;3/2)$]. On the other hand, states \#4, \#5 (with $S=1/2$), and \#6 (with
$S=3/2$) have two electrons in the right well and they are denoted as $(1,2;S)$.
A main feature of this six-state spectrum
in  Fig.\ \ref{spccp}(b) is that, apart from the neighborhoods of the two avoided crossings
(see below Sec.\ \ref{sec:avcr}), the  energy curves for the states \#1, \#2, and \#3 form one
band of parallel lines, whereas the energy curves for the states \#4, \#5, and \#6 form a second
band of parallel lines, and the two bands intersect at two avoided crossings. 
Again the appearance of such three-member bands, grouping together two $S=1/2$ states and one
$S=3/2$ state, is a consequence of the formation of a 3e Wigner molecule (three localized
electrons considering both wells), and this is in consonance with the findings of Ref.\
\cite{yann07.2} regarding the spectrum of three electrons in single anisotropic quantum dots in
variable magnetic fields. 

We further stress that the dominant feature in the spectrum of Fig.\ \ref{spccp}(b) is the small
energy gap between the two $S=1/2$ states \#1 and \#2, which contrasts with the large gap between
the other two $S=1/2$ states \#4 and \#5, a point that will be discussed in detail in Sec.\
\ref{sec:gap} below.  

\subsubsection{Charge densities away from the avoided crossings}
\label{sec:char1}

Further insight into the unique trends and properties of the GaAs double dot with the parameters listed 
in Sec.\ \ref{sec:mbh} is gained through an inspection of the CI-calculated  charge densities, plotted 
in Fig.\ \ref{spccp} for the ground and first five excited states and for both values 
$\kappa=1000$ [see Figs.\ \ref{spccp}(c-h)] and $\kappa=12.5$ [see Figs.\ \ref{spccp}(i-d)] of the 
dielectric constant.

To facilitate the identification and illucidation of the main trends, we display, along with the charge 
densities, the CI-obtained electron occupancies (red lettering) in the left an right wells of the DQD 
(rounded to the second decimal point). [These CI occupations are rounded further to the closest 
integer in order to yield the $n_L$'s and $n_R$'s ($n_L+n_R=3$) used in the notation $(n_L, n_R; S)$ to 
characterize the energy curves in the spectra in Figs.\ \ref{spccp}(a,b)]. Naturally, the charge 
densities are normalized to the total number of electrons $N=3$.

\begin{figure}[t]
\centering\includegraphics[width=8.4cm]{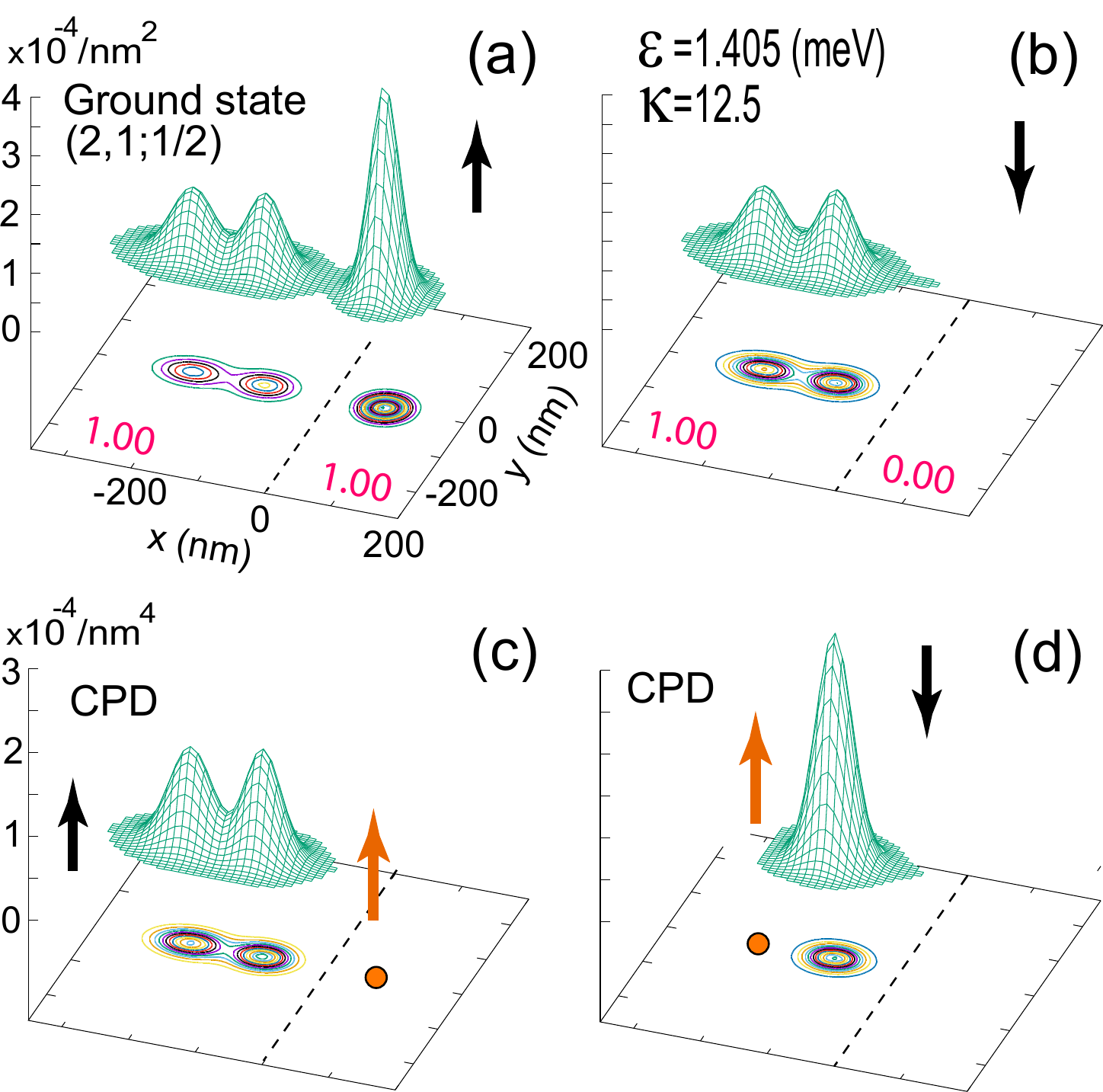}
\caption{
The spin structure of the ground state at $\kappa=12.5$ and $\varepsilon=1.405$ [see Fig.\ 
\ref{spccp}(i) for the corresponding total charge density]. 
(a) The spin-up density. (b) The spin-down density. The red decimal numbers in 
(a) and (b) indicate the CI-calculated left and right occupancies (rounded to the 
second decimal point). (c,d) Examples of two of the associated spin-resolved CPDs. The red dots and 
arrows indicate the position and spin direction of the fixed point. The black arrows indicate the 
spin direction associated with the plotted surface. The spin-resolved densities integrate to the
number of spin-up and spin-down electrons in (a) and (b), respectively. The scale of the vertical 
axes in (c) and (d) is arbitrary, but the same in both cases. 
}
\label{spnstr1}
\end{figure}

\begin{figure}[t]
\centering\includegraphics[width=8.2cm]{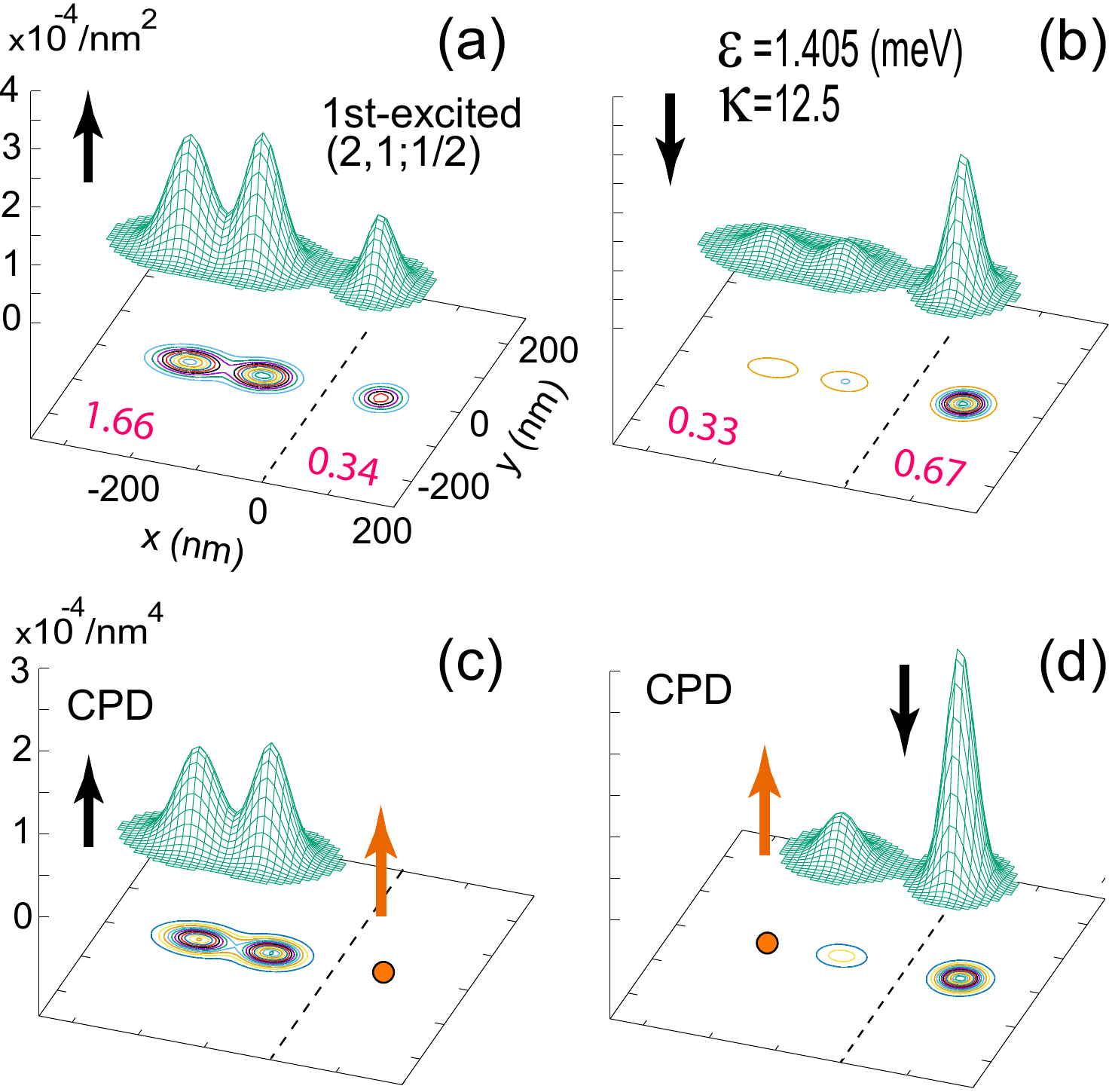}
\caption{
The spin structure of the 1st-excited state at $\kappa=12.5$ and $\varepsilon=1.405$ [see Fig.\ 
\ref{spccp}(j) for the corresponding total charge density]. 
(a) The spin-up density. (b) The spin-down density. 
The red decimal numbers in (a) and (b) indicate the CI-calculated left and right occupancies 
(rounded to the second decimal point). (c,d) Examples of two of the associated spin-resolved CPDs. 
The red dots and 
arrows indicate the position and spin direction of the fixed point. The black arrows indicate the 
spin direction associated with the plotted surface. The spin-resolved densities integrate to the
number of spin-up and spin-down electrons in (a) and (b), respectively. The scale of the vertical 
axes in (c) and (d) is arbitrary, but the same in both cases. 
}
\label{spnstr2}
\end{figure}

Inspection of the charge densities on the left (for $\kappa=1000$, that is, for highly weakened 
inter-electron repulsion) reveals that they conform to
those expected from an independent-particle system. Indeed, in Figs.\ \ref{spccp}(c,f,g), two 
electrons with opposite spins occupy the lowest nodeless $1s$-type single-particle level in the 
right well. At the same time, the single electron in the left well can occupy 
succesively the zero-node [Fig.\ \ref{spccp}(c)], one node [Fig.\ \ref{spccp}(f)], and two-node 
[Fig.\ \ref{spccp}(g)] single-particle states in this well. The 5th-excited state is a $(2,1)$ 
[Fig.\ \ref{spccp}(h)] state with two electrons with opposite spins occupying the lowest nodeless
single-particle level in the left well, whereas the third electron occupies the nodeless
single-particle level of the right well. Finally, the 2nd-excited and 3rd-excited states
[Figs.\ \ref{spccp}(d,e)] are $(0,3)$ states with all three electrons residing in the right well.
In a circular dot, these two states would be degenerate and fully circular with angular momenta
$+1$ and $-1$, but here the degeneracy is lifted because of the influence of the left well and the 
inter-well neck.

The charge densities on the right (for $\kappa=12.5$, case of GaAs) deviate strongly from those
expected from an independent-particle system. Indeed the formation of a strong 2e WM in the left
well and of a weaker 2e WM in the right well is clearly seen. For example, contrast
the $(2,1)$ independent-particle density in Fig.\ \ref{spccp}(h) with the $(2,1)$ WM density
in Figs.\ \ref{spccp}(i,j,k) and the $(1,2)$ independent-particle density in Fig.\ \ref{spccp}(c)
with the $(1,2)$ WM density in Figs.\ \ref{spccp}(l,m,n). We also note the absence of states with higher 
than double-occupancy in any of the DQD wells, unlike the findings for the case of highly quenched
inter-electron interaction discussed above; such states may emerge as highy-excited states for GaAs DQDs
with the parameters used here (as well as in the experiments \cite{kim21}), and need not to be of concern
for future application of such DQDs. 

\subsubsection{Spin structure away from the avoided crossings}
\label{sec:spnstr}

The charge densities of the states \#1, \#2, and \#3 in the three-member band for $\kappa=12.5$
[see Figs.\ \ref{spccp}(i,j,k)] are very similar. However the corresponding spin structures are
different, as shown explicitly below. Differences between the spin structures may be investigated with 
the employment of: (1) spin-resolved densities, and (2) spin-resolved conditional 
probability disctributions. The concept of spin-resolved densities is self-evident. The 
concept of the spin-resolved CPDs is more complicated, and the full definition is provided in 
Appendix \ref{spdcpd}. At the intuitive level, the spin-resolved CPD addresses the following question:
given the specific (fixed) location ${\bf r}_0$ of an electron with a definite spin (up or down), what is
the probability distribution at another location ${\bf r}$ for finding another electron with a definite 
(up or down) spin? 

As an example of the spin-structure analysis that can be achieved with the CI calculations, we 
analyze below the two cases of the ground state and the 1st-excited state for $\kappa=12.5$ (GaAs) 
and $\varepsilon=1.405$ meV.

Fig.\ \ref{spnstr1}(a) and Fig.\ \ref{spnstr1}(b) display, respectively, the spin-up and spin-down 
densities for the ground state mentioned above; compare Fig.\ \ref{spccp}(i) for the total charge 
density. From these two spin-resolved densities, it is evident that the spin structure of this 
ground state conforms to the following familiar expression \cite{vinc00,burk17,copp12,kim21}
in the theory of three-electron qubits:
\begin{align}
(|duu\rangle-|udu\rangle)/\sqrt{2},
\label{spin1}
\end{align}
where $u$ and $d$ denote an up and down spin, respectively, with the three spins arranged in a line 
from left to right in three ordered sites, 1, 2, and 3.

Further confirmation for the spin structure in Eq.\ (\ref{spin1}) is obtained from the 
spin-resolved CPDs, two examples of which are displayed in Figs.\ \ref{spnstr1}(c) and 
\ref{spnstr1}(d). Indeed both the spin kets in Eq.\ (\ref{spin1}) are compatible with a first
electron with spin up being fixed at the third site. Then the first spin ket in Eq.\ (\ref{spin1})
allows the presence of a spin-up electron in its second site with probability 1/4, whereas
the second spin ket allows the presence of a spin-up electron in its first site, again with 
equal probability 1/4. This is clearly in agreement with the CI CPD plotted in Fig.\ 
\ref{spnstr1}(c), which exhibits two humps of similar height at sites No.\ 1 and No.\ 2. 
Likewise, only the second spin ket in Eq.\ (\ref{spin1}) is compatible with a first electron with 
spin up being fixed in the first site, and this spin ket allows the presence of a spin-down 
electron in its second site with probability 1/4. This again is in agreement with the CI CPD
plotted in Fig.\ \ref{spnstr1}(d), which exhibits a single hump at site No.\ 2.

Fig.\ \ref{spnstr2}(a) and Fig.\ \ref{spnstr2}(b) display the spin-up and spin-down densities for
the associated 1st-excited state; compare Fig.\ \ref{spccp}(j) for the total charge density. From
these two spin-resolved densities, one can conclude that the spin structure of this 1st-excited
state conforms to a second familiar expression \cite{vinc00,burk17,copp12,kim21} in the theory
of three-electron qubits, namely
\begin{align}
(2|uud\rangle-|duu\rangle-|udu\rangle)/\sqrt{6}.
\label{spin2}
\end{align}

Indeed, from Eq.\ (\ref{spin2}), one can derive that the expected spin-up occupancy for the most 
leftward and middle positions of the three spins is $5/6$ in both cases, 
yielding $5/3=1.666$ for the expected spin-up occupancy in the left dot, 
in agreement with the CI value of 1.66 highlighted in red in Fig.\ \ref{spnstr2}(a).
Similarly the expected spin-up occupancy for the right dot from Eq.\ (\ref{spin2}) is $1/3=0.333$,
in agreement with the CI-value of 0.34 highlighted in red in Fig.\ \ref{spnstr2}(a).
Moreover from Eq.\ (\ref{spin2}), one can derive that the expected spin-down occupancy for the most
leftward and middle positions of the three spins is $1/6$ in both cases, 
yielding $1/3=0.333$ for the expected spin-down occupancy in the left dot, in agreement with the CI 
value of 0.33 highlighted in red in Fig.\ \ref{spnstr2}(b).
Finally the expected spin-down occupancy for the right dot from Eq.\ (\ref{spin2}) is $2/3=0.666$,
in agreement with the CI-value of 0.67 highlighted in red in Fig.\ \ref{spnstr2}(b).

As aformentioned, further confirmation for the spin structure in Eq.\ (\ref{spin2}) 
can be obtained from the spin-resolved CPDs, two examples 
of which are displayed in Figs.\ \ref{spnstr2}(c) and 
\ref{spnstr2}(d). Indeed only the second and third spin kets in Eq.\ (\ref{spin2}) are compatible 
with a first electron with spin up being fixed at the third site. Then the second spin ket in Eq.\ 
(\ref{spin2}) allows the presence of a spin-up electron in its second site with probability 1/6, 
whereas the third spin ket allows the presence of a spin-up electron in its first site, again with 
equal probability 1/6. This is clearly in agreement with the CI CPD plotted in Fig.\ 
\ref{spnstr2}(c), which exhibits two humps of similar height at sites No.\ 1 and No.\ 2. 
Likewise, only the first and third spin kets in Eq.\ (\ref{spin2}) are compatible with a first 
electron with spin up being fixed in the first site. Then the first spin ket allows the presence of 
a spin-down electron in its third site with probability 2/3, whereas the third spin ket allows 
the presence of a spin-down electron in its second site with probability 1/6. This again is in 
agreement with the CI CPD plotted in Fig.\ \ref{spnstr2}(d), which exhibits
two humps with relative height 1/4 at sites No.\ 2 and No. 3. 

\textcolor{black}{
We stress that the two expressions in Eqs.\ (\ref{spin1})-(\ref{spin2}) for the spin structures of
three fermions are not the most general ones. The most general \cite{note5} expression for a total spin 
$S=1/2$ with a total-spin projection $S_z=1/2$ is \cite{suzubook}:
\begin{align}
\begin{split}
\sqrt{\frac{2}{3}} \sin\vartheta |uud\rangle & +
\left( \sqrt{\frac{1}{2}} \cos\vartheta - \sqrt{\frac{1}{6}} \sin\vartheta \right) |udu\rangle \\
& - \left( \sqrt{\frac{1}{2}} \cos\vartheta + \sqrt{\frac{1}{6}} \sin\vartheta \right) |duu\rangle,
\end{split}
\label{spingen}  
\end{align} 
where the angle $\vartheta$ can take any value in the interval $[\pi/2,3\pi/2]$. 
Eq.\ (\ref{spin1}) is recovered for $\vartheta=\pi$, whereas its companion orthogonal expression 
(\ref{spin2}) is recovered for $\vartheta=\pi/2$. Thus it is gratifying to see that the FCI solutions, 
analyzed with the CPDs, that are associated with the hybrid-qubit device of Ref.\ \cite{kim21} do 
not deviate from the spin structures invoked in building the theorical models of three-electron qubits 
\cite{vinc00,burk17,copp12}. 
}

\begin{figure}[t]
\centering\includegraphics[width=7.5cm]{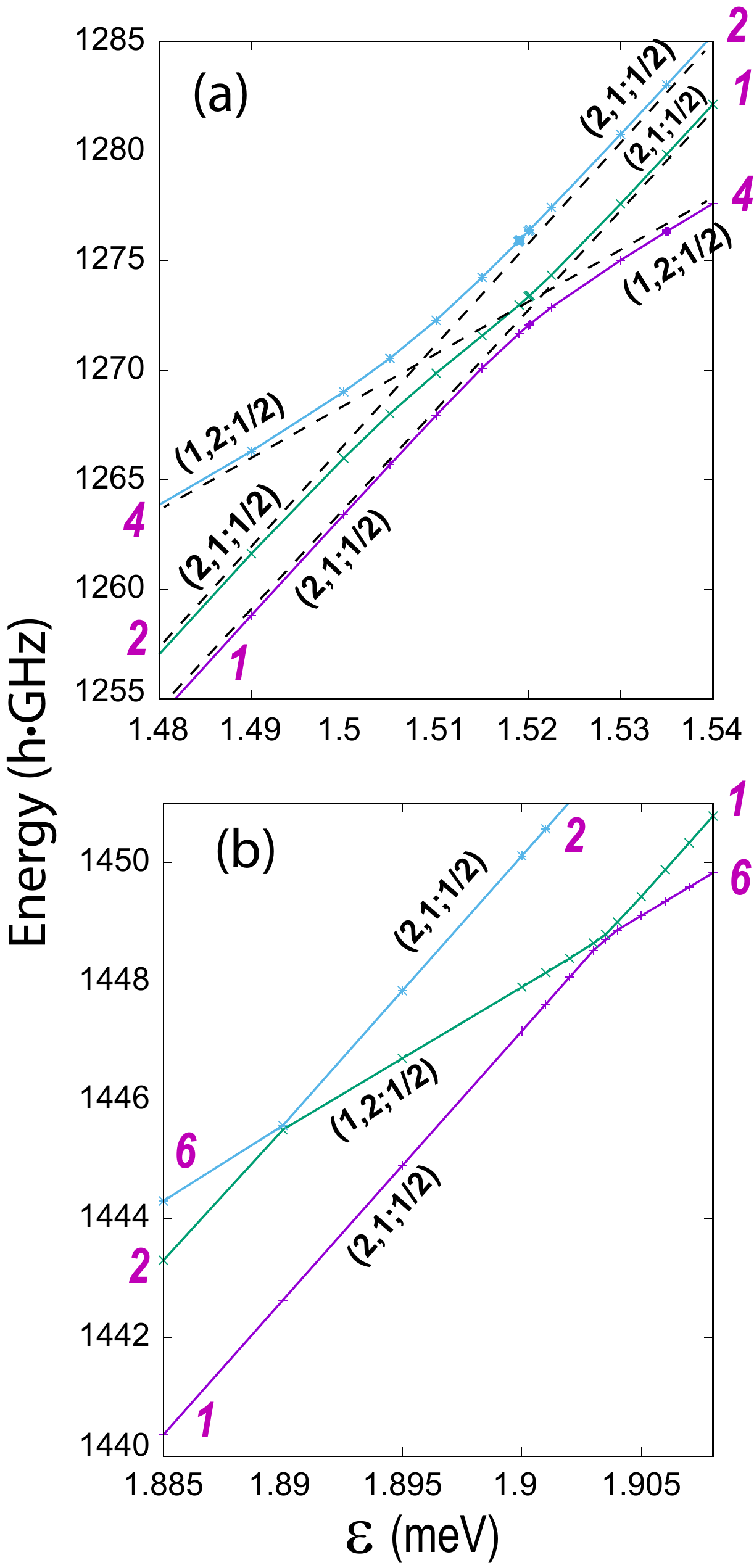}
\caption{
Magnification of the neighborhoods of the CI avoided crossings appearing in Fig.\ \ref{spccp}(b)
($\kappa=12.5$, case of GaAs). Only the $S=1/2$ states, relevant to the hybrid qubit, are shown.
(a) The left avoided crossing in the neighborhood of 1.49 meV $<~\varepsilon~<$ 1.54 meV.
(b) The right avoided crossing in the neighborhood of 1.885 meV $<~\varepsilon~<$ 1.908 meV.
}
\label{spcacs}
\end{figure}

\subsection{The avoided crossings}
\label{sec:avcr}

In Fig.\ \ref{spcacs}(a) and Fig.\ \ref{spcacs}(b), we display magnifications of the neighborhoods
of the left and right CI avoided crossings, respectively, appearing in the spectrum of the GaAs
double dot [Fig.\ \ref{spccp}(b)]. Only the $S=1/2$ states are shown, because the $S=3/2$ states
are not relevant for the workings of the hybrid qubit \cite{kim21,copp12,vinc00,burk17}. 

The left avoided crossing (situated in the neighborhood of 1.49 meV $<~\varepsilon~<$ 1.54 meV)
is formed through the interaction of the three curves \#1, \#2, and \#4 [we keep the same numbering
of curves here as in Fig.\ \ref{spccp}(b)]. On the other hand, the curves \#1, \#2, and \#6
participate in the formation of the right avoided crossing in the neighborhood of 1.885 meV
$<~\varepsilon~<$ 1.908 meV. We note that, according to the FCI calculation, the two avoided crossings
are separated by a detuning energy-interval of $\sim 400$ $\mu$eV, which agrees with the experimentally
determined value for the hybrid qubit device in Ref.\ \cite{kim21}.

The continuous lines in both panels of Fig.\ \ref{spcacs} represent the socalled {\it adiabatic\/}
paths, which the system follows for slow time variations of the detuning. For fast time variations of
the detuning, or with an applied laser pulse, the system can instead follow the {\it diabatic\/} paths
indicated explicitly with dashed lines in Fig.\ \ref{spcacs}(a) and thus jump from one adiabatic line to
another; this occurs according to the celebrated Landau-Zener-St\"{u}ckelberg-Majorana
\cite{cao13,copp12.2,burk13} dynamical interference theory.

We further mention that the position and the asymmetric anatomy of the two avoided crossings play an
essential role in the operation of the hybrid qubit. Indeed, the qubit is initialized on line \#4 and
in a $(1,2;1/2)$ ground-state configuration at a value of detuning far to the right of the left
crossing. Then by decreasing the magnitude of the detuning in an adiabatically evolving manner, the state
of the qubit moves along the \#4 line and is brought in the neighborhood of the left avoided crossing,
where a laser pulse induces a small-gap-enabled diabatic transition to the line \#1 [a $(2,1;1/2)$ line].
Next by increasing the detuning value, the qubit operation cycle proceeds by moving adiabatically
backwards along the \#1 line and through the right avoided crossing transitioning to the line \#6
[a $(1,2;1/2)$ line], where the readout can be implemented, aided by the large energy gap between the two
$(1,2;1/2)$ states \#4 and \#6 \cite{kim21,copp14.2}.

\begin{figure}[t]
\centering\includegraphics[width=8.2cm]{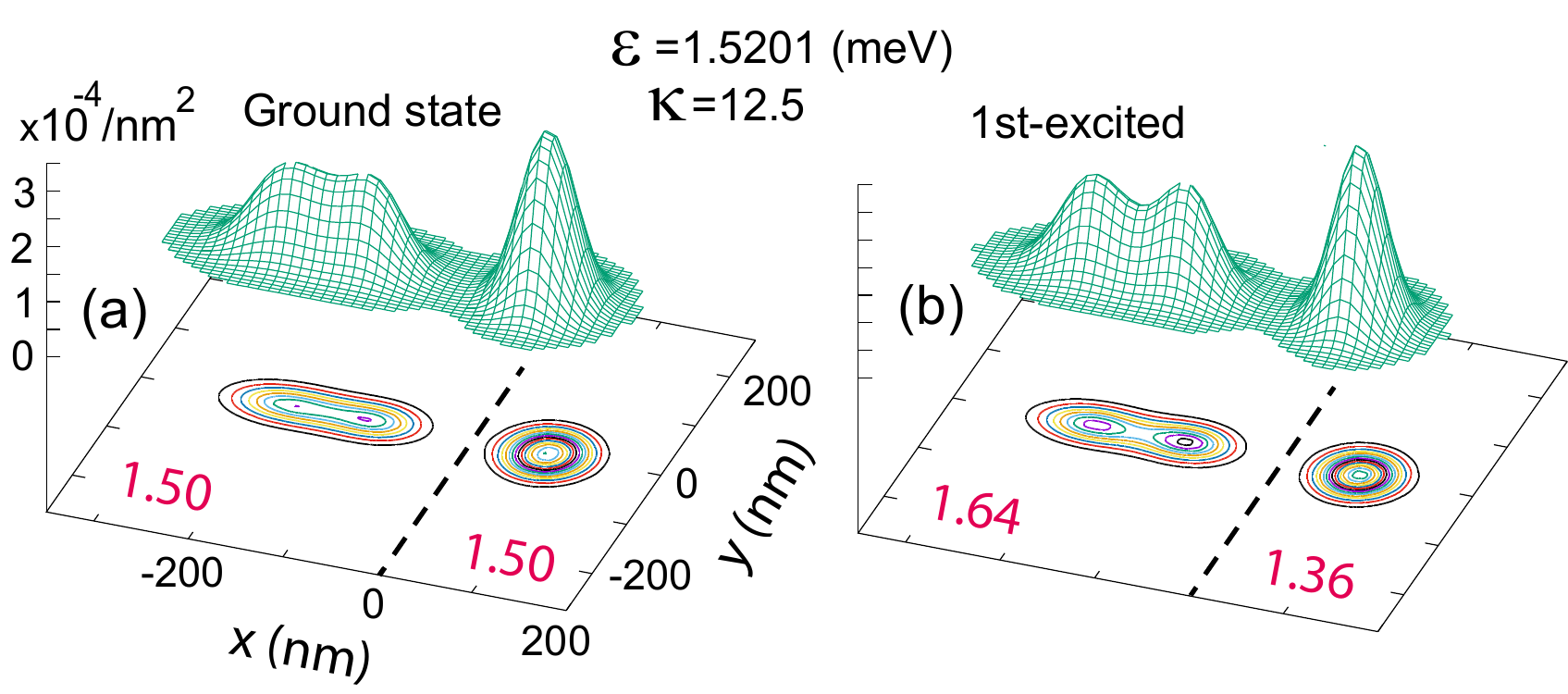}
\caption{
Examples of charge densities at the left avoided crossing of the spectrum in Fig.\ \ref{spccp}(b)], 
namely at $\varepsilon=1.5201$ and $\kappa=12.5$ (a) The ground state. (b) The 1st-excited state. Both
states have a total spin $S=1/2$. The red decimal numbers indicate the left-right electron occupancies
according to the CI calculation (rounded to the second decimal point). 
}
\label{dvcr}
\end{figure}

\subsubsection{Charge densities at the avoided crossings: Mixing of WM states}
\label{sec:mix}

Inside the neighborhood of the avoided crossings, it is expected that the ground state will be a 
superposition of the two different wave functions $(2,1)$  (with two electrons in the left well) and
$(1,2)$ (with two electrons in the right well). For $\kappa=12.5$, this is confirmed by the CI 
ground-state charge density [see Fig.\ \ref{dvcr}(a)] at the detuning value of $\varepsilon=1.5201$ meV 
[middle point in the neighborhood of the left avoided crossing; see Fig.\ \ref{spcacs}(a)]. Indeed the 
associated left and right electron occupancies equal both 1.50, suggesting the the CI ground state is a 
superposition given by $(|1\rangle + |4\rangle)/\sqrt{2}$. 
Likewise, at the same value of $\varepsilon=1.5201$ meV, the charge density of the 1st-excited CI state 
[see Fig.\ \ref{dvcr}(b)] exhibits left and right electron occupancies equal to 1.64 and 1.36,
respectively, suggesting that this state is a superposition given by $0.8|1\rangle - 0.6 |4\rangle$. 

\begin{figure}[t]
\centering\includegraphics[width=7.5cm]{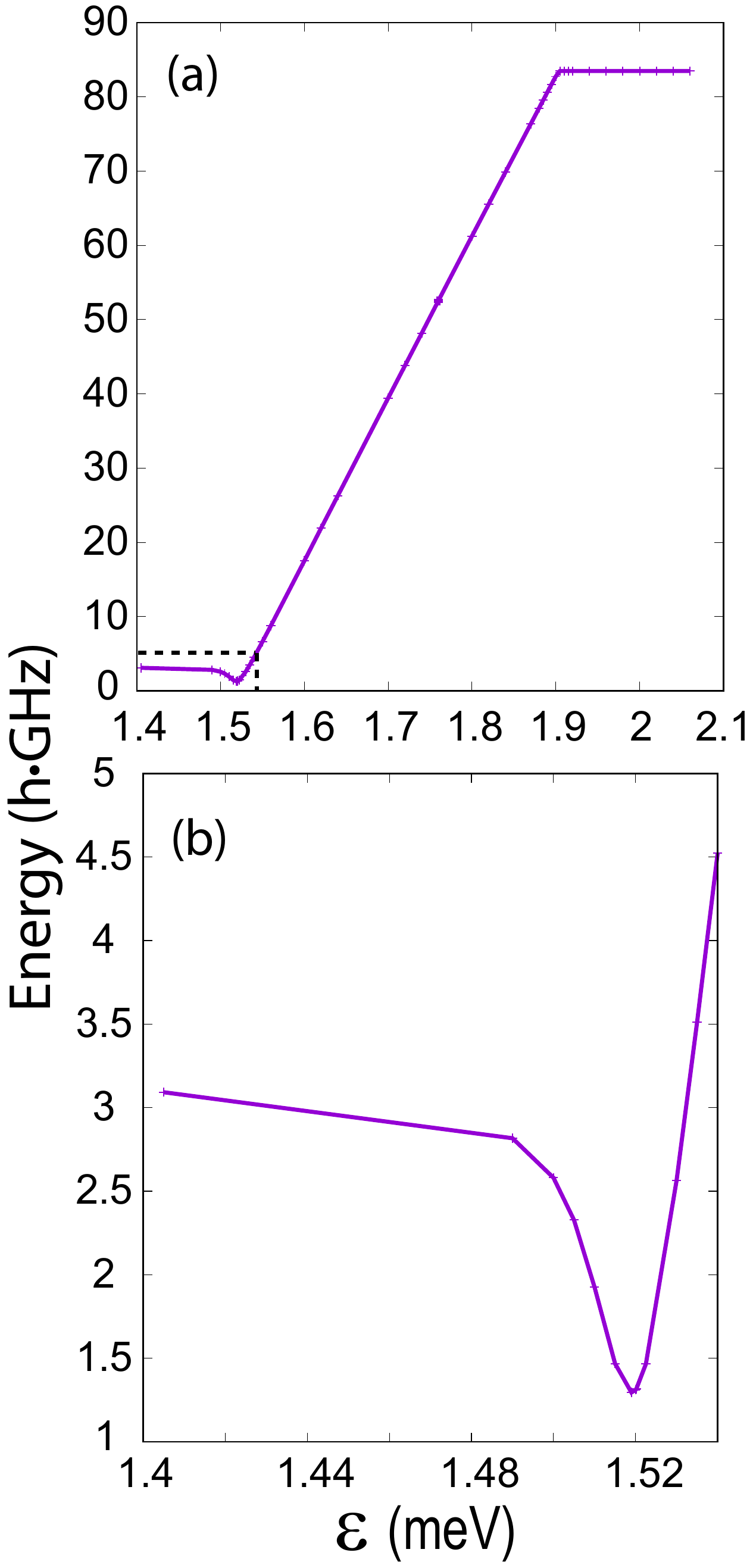}
\caption{
The CI-calculated energy gap between ground and 1st-excited states as a function of
the detuning for $\kappa=12.5$ (GaAs) [see the spectrum in Fig.\ \ref{spccp}(b)]. (a) In the 
broader range from 1.4 meV to 2.1 meV covering both avoided crossings. (b) In a smaller
range [see the dashed-border box in (a)] covering the left avoided crossing.
}
\label{spce1me0}
\end{figure}

\begin{figure*}
\centering\includegraphics[width=15cm]{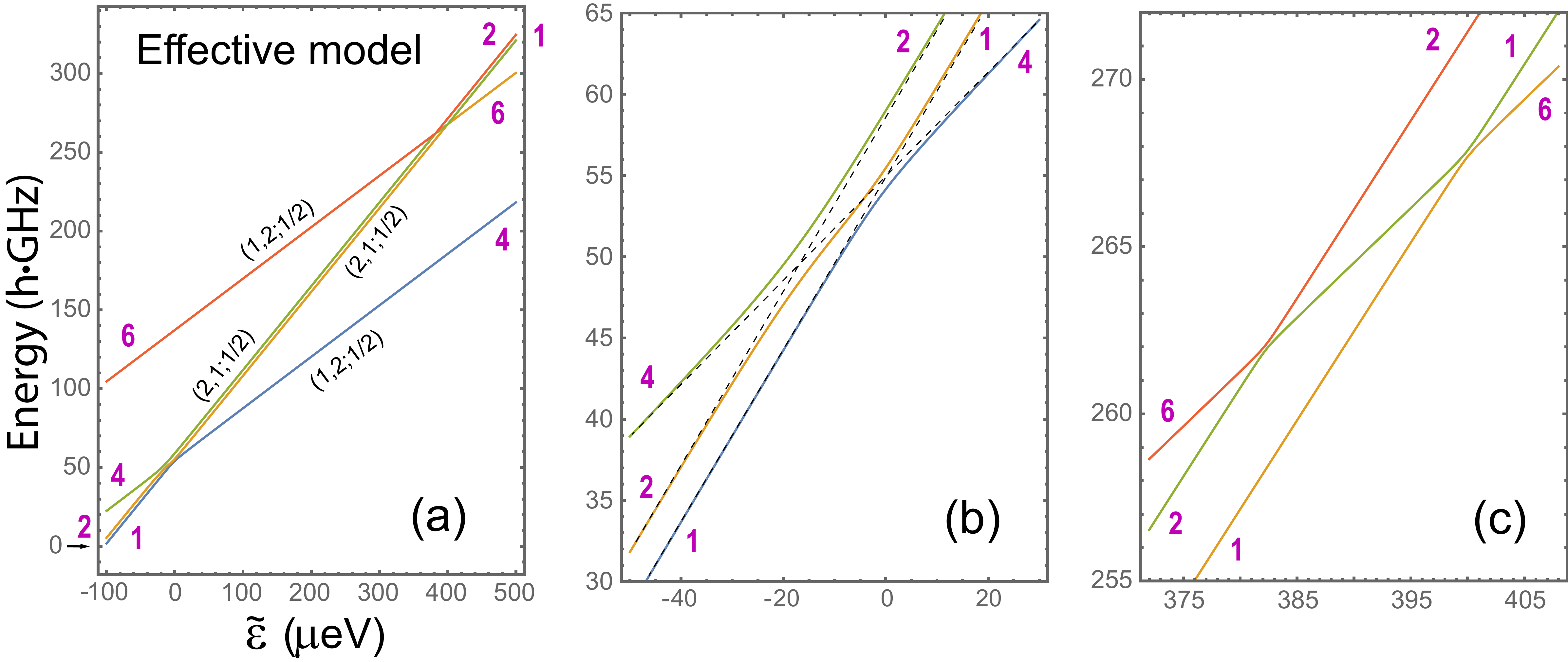}
\caption{
The spectrum of the four hybrid-qubit states 
$|1\rangle$, $|2\rangle$, $|4\rangle$, and $|6\rangle$ (with $S=1/2$)
as a function of the detuning parameter $\widetilde{\varepsilon}=\varepsilon - \varepsilon_0$, 
calculated with the effective matrix Hamiltonian in Eq.\ (\ref{eqtoyham});
$\varepsilon_0=1.51$ meV. (a) The full spectrum. (b) Magnification of the left avoided crossing. 
(c) Magnification of the right avoided crossing. 
For the employed values of the 8 parameters $c_L$, $\delta E_L$, $c_R$, $\delta E_R$, $\delta_1$, 
$\delta_2$, $\delta_3$, $\delta_4$, see the text. Note the remarkable agreement between the 
4-state effective spectrum here and the corresponding one of the 4 lowest $S=1/2$ states in Figs.\ 
\ref{spccp}(b) and \ref{spcacs}, which was generated from the FCI calculations. 
}
\label{spctoy}
\end{figure*}

\begin{figure}[t]
\centering\includegraphics[width=8.4cm]{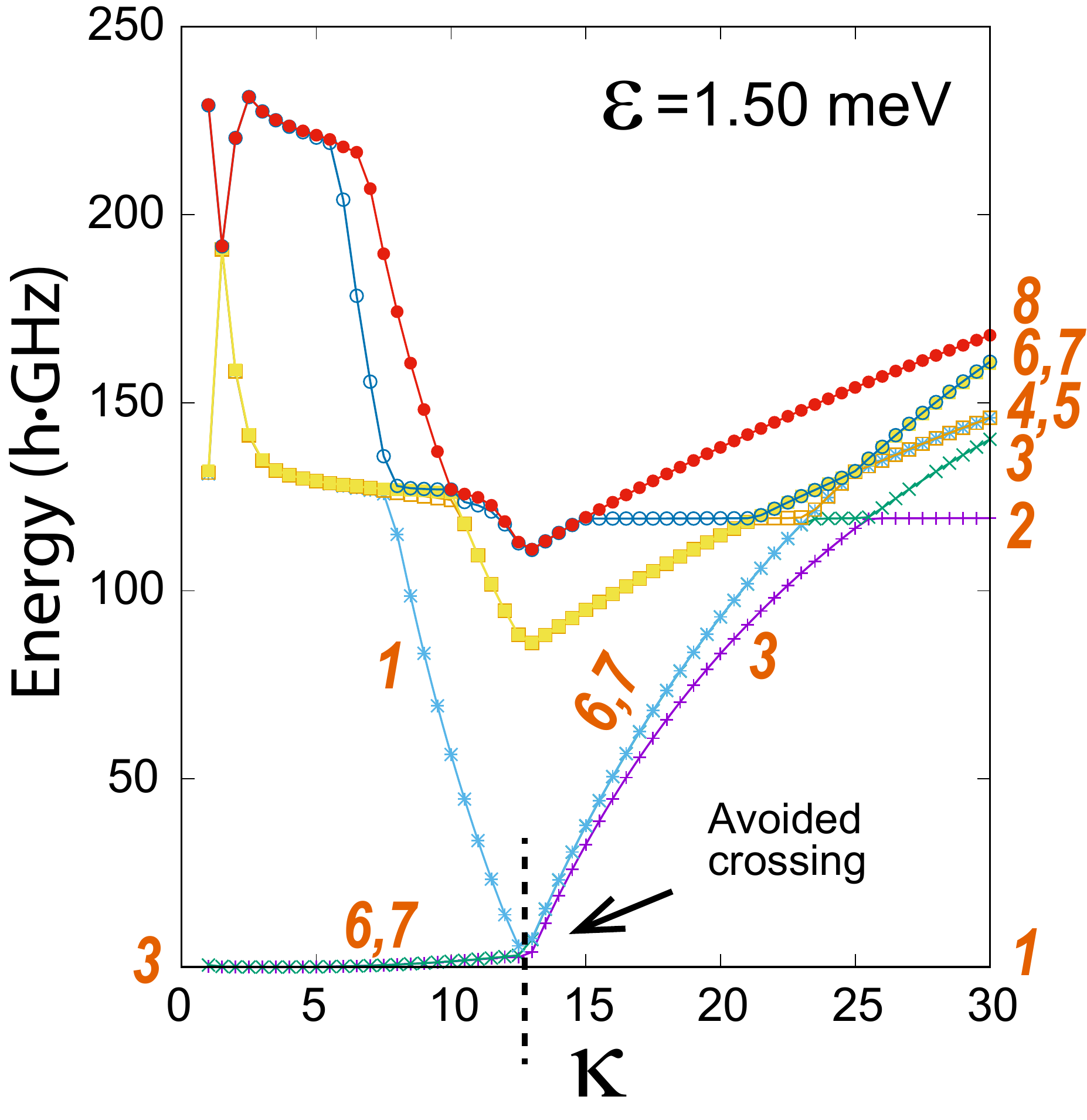}
\caption{
Spectrum of the double dot as a function of the dielectric constant $\kappa$ at a detuning value 
of $\varepsilon=1.50$ meV. The numbering of the lowest 8 states is performed at $\kappa=30$, and
this numbering is maintained for the whole $\kappa$ range plotted. The vertical dashed line at
$\kappa=12.7$ indicates the neighborhood of the avoided crossing. Further to the left of the avoided 
crossing, the ground state is state \#3 of $(2,1;1/2)$ character; further to the right of the avoided 
crossing, the ground state is state \#1 of $(1,2;1/2)$ character. In the neighborhood of the
avoided crossing, the ground state is a mixed state. In this plot, the ground-state 
energies were taken at all instances to coincide with the zero of the energy scale. 
For the $(n_L,n_R;S)$ designations, see Fig.\ \ref{spckd}.
}
\label{spck}
\end{figure}

\subsection{The energy gap between the ground and 1st-excited states}
\label{sec:gap}

To further highlight the notable specifications of the GaAs double dot considered in this paper, we
display in Fig.\ \ref{spce1me0} the CI-calculated energy gap between the ground and the 1st-excited state
as a function of the detuning for $\kappa=12.5$ (GaAs) [see the spectrum in Fig.\ \ref{spccp}(b)]. Fig.\
\ref{spce1me0}(a) displays the broader view in the detuning range from 1.4 meV to 2.1 meV covering
both avoided crossings. In reference to the spectrum in Fig.\ \ref{spccp}(b), this energy gap
involves the following states: (1) The states \#1 and \#2 to the left of the left avoided crossing,
(2) the states \#4 and \#1 between the two avoided crossings, and (3) the states \#4 and \#6 to the
right of the right avoided crossing. Fig.\ \ref{spce1me0}(a) highlights the overarching rise by an order
of magnitude of this energy gap, from $\sim 3$ h$\cdot$GHz to $\sim 83$ h$\cdot$GHz, as the spectrum
transitions from the $(2,1)$ to the $(1,2)$ configurations with two electrons in the left and right dots,
respectively. We note that this behavior is in excellent agreement with the sharp rise of the
corresponding energy gap within a detuning range of $\sim 400$ $\mu$eV proposed in the experimental paper
of Ref.\ \cite{kim21} [see Fig.\ S5(b) therein].

Moreover, Fig.\ \ref{spce1me0}(b) magnifies the corresponding energy gap in a smaller range [see the
dashed-border box in Fig.\ \ref{spce1me0}(a)] covering only the left avoided crossing. The CI-calculated
curve in \ref{spce1me0}(b) exhibits remarkable overall agreement with the corresponding measured one in
Fig.\ 1(b) of Ref.\ \cite{kim21} [see also Fig.\ S5(b) therein].

\begin{figure*}
\centering\includegraphics[width=17cm]{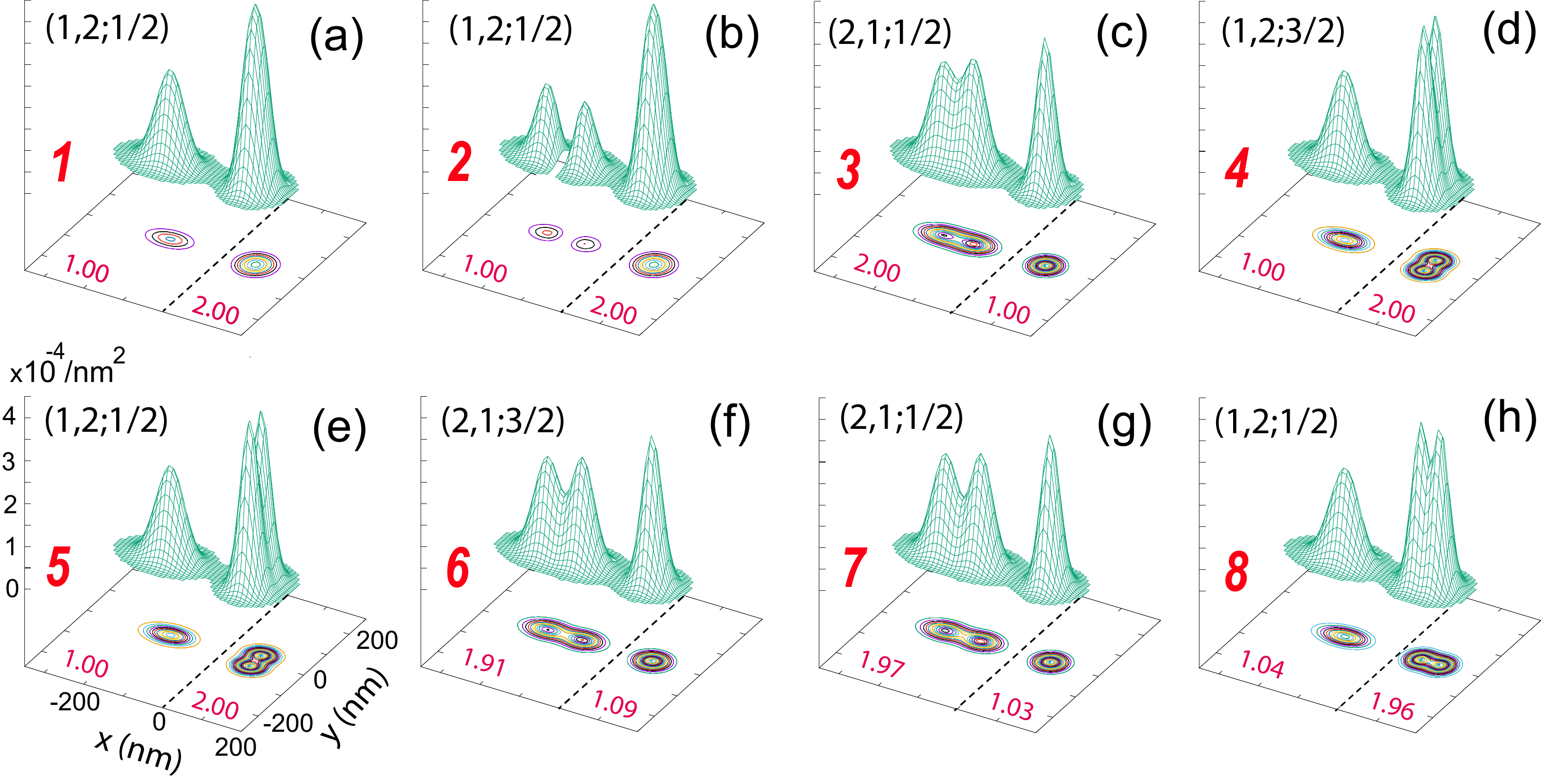}
\caption{
Charge densities for the lowest 8 states of the double quantum dot at $\kappa=30$ and $\varepsilon=1.50$
meV. The numbering of states is according to Fig.\ \ref{spck}. The red decimal numbers indicate the 
left-right electron occupancies according to the CI calculations (rounded to the second decimal point). 
For a detailed description, see the text.
}
\label{spckd}
\end{figure*}

\subsection{The effective matrix Hamiltonian}
\label{sec:effham}

In this section, we extract from the CI spectra the phenomenological effective matrix Hamiltonian 
\cite{kim21,copp14.2} that has played a central role in the experimental dynamical control of the 
hybrid cubit. The general form of this 4$\times$4 matrix Hamiltonian is:
\begin{align}
H_M=\left(
\begin{array}{cccc}
c_L \widetilde\varepsilon/2 & 0 & \delta_1 & -\delta_2 \\
0 & c_L \widetilde\varepsilon/2 + \Delta E_L & -\delta_3 & \delta_4 \\
\delta_1 & -\delta_3 & c_R \widetilde\varepsilon/2 & 0\\
-\delta_2 & \delta_4 & 0 & c_R \widetilde\varepsilon/2+\Delta E_R
\end{array}
\right), 
\label{eqtoyham}
\end{align}
where $\widetilde\varepsilon=\varepsilon-\varepsilon_0$ denotes a renormalized interdot detuning, and
the other elements of the matrix follow the notation used in Ref.\ \cite{copp14.2} (see the Theory section 
therein). 

A good fit (see Fig.\ \ref{spctoy}) with the CI spectrum in Figs.\ \ref{spccp}(b) and \ref{spcacs} is 
achieved by setting $c_L=4.4$, $\Delta E_L=15$ $\mu$eV, $c_R=2.7$, $\Delta E_R=340$ $\mu$eV, 
$\delta_1=0.657$ $\mu$eV, $\delta_2=0.090$ $\mu$eV, $\delta_3=1.207$ $\mu$eV, $\delta_4=0.075$ $\mu$eV, 
and $\varepsilon_0=1.50$ meV.

The effective matrix Hamiltonian in Eq.\ (\ref{eqtoyham}) reflects (within the plotted 
window) two properties of the FCI spectrum in Fig.\ \ref{spccp}(b) that are instrumental 
(see, e.g., \cite{cao17.2,copp14.2}) for the successful operation of the hybrid qubit, 
namely, the quasi-linear dependence of $H_M$ on the detuning $\varepsilon$ and the quasi-parallel
behavior of both the two $(2,1)$ states (states \#1 and \#2) and the two $(1,2)$ states 
(states \#4 and \#6). We note a difference between Refs.\ \cite{kim21,copp14.2} and
the CI result for $H_M$. Namely, Refs.\ \cite{kim21,copp14.2} assume the values $c_L=1$ and $c_R=-1$
associated with 45$^o$ and -45$^o$ slopes of the associated lines, respectively, while the CI 
result produces different slopes associated with  $c_L=4.4$ and $c_R=2.7$. 
\textcolor{black}{We note, however, that the topology of the energy spectrum remains unaltered.}

We further note that there are several derivations \cite{copp12,ferr14} of the effective Hamiltonian in 
Eq.\ (\ref{eqtoyham}) starting from approximate many-body Hamiltonians that include the interaction at 
the level of a two-site (left and right well) Hubbard-type modeling. These derivations involve several 
additional qualitative approximations and are not applicable in the case of strong $e-e$ correlations and
Wigner-molecule formation (see Ref.\ \cite{note3}). Thus the reaffirmation demonstrated above regarding 
the overall structure of the phenomenological-in-nature effective matrix Hamiltonian [Eq.\ 
(\ref{eqtoyham})], achieved here through the use of FCI-based {\it ab-initio} calculations carried out 
in the regime of strong correlations and Wigner-molecule formation, is a notable result.   

\section{CI spectra as a function of the dielectric constant}
\label{spectdie}

Further insights into the effects of the interelectron interaction can be gained by an inspection
of the CI spectra as a function of the dielectric constant $\kappa$. Fig.\ \ref{spck} portrays the
spectrum of the double dot as a function of the dielectric constant $\kappa$ at a detuning value 
of $\varepsilon=1.50$ meV. The numbering of the lowest 8 states is performed at $\kappa=30$, and
this numbering is maintained for the whole $\kappa$ range plotted. The vertical dashed line at
$\kappa=12.7$ indicates the neighborhood of the avoided crossing. Further to the left of the avoided 
crossing, the ground state is state \#3 of $(2,1;1/2)$ character; further to the right of the avoided 
crossing, the ground state is state \#1 of $(1,2;1/2)$ character. In the neighborhood of the avoided 
crossing, the ground state is a mixed state. In this plot, the ground-state energies were taken at all 
instances to coincide with the zero of the energy scale. The $(n_L,n_R;S)$ designations are not plotted 
in this figure, but they can be traced through Fig.\ \ref{spckd}, which portrays the associated 
charge densities.

From the charge densities in Fig.\ \ref{spckd}, one sees that at $\kappa=30$ the three excited states 
\#3, \#6, and \#7 display a 2e WM in the left dot, with the WM aligned along the $x$-axis.
With decreasing $\kappa$ (increasing Coulomb
repulsion), the energies of this triad of WM states exhibit a sharp drop in magnitude, and as a result
they become the lowest-energy band for $\kappa< 12.7$, a fact which agrees with the findings in Sec.\ 
\ref{sec:big}. In addition the \#6 ($S=3/2$) and \#7 ($S=1/2$) curves become quasi-degenerate, and they 
exhbit a small energy gap ($\leq 3$ h$\cdot$GHz) from the ground state (\#3 state), again in agreement 
with the findings in Sec.\ \ref{sec:big}; see curves numbered \#1, \#2, and \#3 in Fig.\ \ref{spccp}(b). 

Of interest is the fact that the ground state at $\kappa=30$ (\#1 state) exhibits a charge density that
can be undestood purely with the help of three non-interacting electrons [see Fig.\ \ref{spckd}(a)], 
namely one spin-up and one spin-down electrons occupying the nodeless 1$s$ lowest single-particle state 
of the circular right well and one spin-up electron occupying the nodeless $1s_x1s_y$ lowest 
single-particle state in the asymmetric left well. Naturally, as $\kappa$ decreases, the intrinsic 
structure of the associated wave function transitions smoothly to that of a weak 2e WM in the right dot 
aligned along the $y$-axis, namely the charge density in Fig.\ \ref{spckd}(a) transitions to that 
portrayed in Fig.\ \ref{spccp}(l).

Of interest also are the intrinsic structures (at $\kappa=30$) of the remaining four excited states,
\#2, \#4, \#5, and \#8; they are a testament to the variety and complexity of the 3e DQD system. Indeed,
from Fig.\ \ref{spckd}(b), one sees that excited state \#2 is a non-interacting 3e state similar to state
\#1, but with the single spin-up electron in the left well promoted to the one-node $1p_x1s_y$ 
single-particle state. On the other hand, the charge densities in Figs.\ \ref{spckd}(d) and \ref{spckd}(e) 
demonstrate that both states \#4 and \#5 exhibit a 2e WM in the right well, with the WM aligned along
the $y$-axis. Finally, the charge density in Fig.\ \ref{spckd}(h) reveals the formation of a 2e WM
in the right well, but with the WM aligned along the $x$-axis.

\section{Summary}
\label{sec:sum}

We presented extensive FCI numerical results that addressed both the energetics and the intrinsic 
structure of the many-body wave functions through the calculation of charge and spin-resolved densities,
as well as spin-resolved conditional probability distributions (spin-resolved two-body correlation
functions). Going beyond the two-particle WMs studied with CI treatments in a single dot 
\cite{erca21,erca21.2,kim21,urie21}, this paper advances and enables microscopic investigations
of key features appearing in the low-energy spectrum as a function of detuning of actual experimentally 
fabricated GaAs three-electron asymmetric HDQD qubits. These features include  the strong suppression of 
level gaps compared to the non-interacting-electrons case and the appearance of a pair of avoided crossings 
between triad of levels corresponding to different electron occupancies in the left and right wells. 
Treating the HDQD as an integral unit, we further showed in depth that these features arise from the 
emergence of WMs. Away from the avoided crossings, it was found that the WMs can be associated with 
molecular configurations centered in the individual wells. In the neighborhood of the avoided crossings, 
the WMs interact and form more complex resonating structures.  

Earlier experimental observations had suggested that the qubit's spectral features could be codified using 
simple phenomenological effective matrix Hamiltonians [see, e.g., Eq.\ (\ref{eqtoyham})]. Our double-dot 
extensive calculations, encompassing both energetic and structural aspects, enabled tracing  of the 
microscopic origins of such phenomenological treatments. This phenomenology has been identified here to 
emerge from the complex nature of the many-body problem encountered due to the strong e-e correlations, 
leading to electron localization and formation of Wigner molecules. 

Previous tentative derivations \cite{copp12,ferr14} of matrix Hamiltonians [of a similar structure as
in Eq.\ (\ref{eqtoyham})], starting from approximate two-site Hubbard-type modeling, involve qualitative 
approximations and are not applicable in the case of WM formation. Consequently, the present CI-based 
derivation of the effective matrix Hamiltonian in Eq.\ (\ref{eqtoyham}), achieved here via analysis using 
the results of FCI calculations that account fully for strong-correlation effects within each well and 
WM formation, is an unexpected auspicious result.  

\textcolor{black}{
Our multi-dot FCI method can be expanded to incorporate the valley degree of freedom as an isospin in full 
analogy with the usual spin. Such an expansion will enable the acquisition of numerical results complete 
with full spin-isospin assignments that will reveal the underlying $SU(4) \supset SU(2) \times SU(2)$ 
group-chain organization of the spectra of Si double-quantum-dot qubits. This valleytronic CI \cite{yann22}
will be a most effective tool for analysing the spectra of qubits and for providing effective matrix 
Hamiltonians that differentiate between cases when the first excited state belongs to the same or different 
valleys, as a result of the competition \cite{erca21,erca21.2} between the valley gap and strong $e-e$ 
interactions. In this respect, we note the case of a 5e-HDQD Si/SiGe qubit where more complex spectra, 
requiring an $n \times n$ matrix with $n>4$, have been recently experimentally discovered \cite{note6}. 
An application of the valleytronic CI to two-qubit gates appears to be feasible \cite{yann22} 
in the near future, based on our estimates of the size of the required Hilbert spaces for an ensemble of 
$N \leq 6$ electrons confined in two interacting DQDs.}
Finally, beyond the GaAs-based devices, our method 
can be directly applied to similar devices built from other 
one-band materials, like holes in Germanium \cite{veld20,veld21} or Silicon \cite{urie21} qubits. 

\section{acknowledgments}

This work has been supported by a grant from the Air Force Office of Scientific Research (AFOSR) 
under Award No. FA9550-21-1-0198. Calculations were carried out at the GATECH Center for 
Computational Materials Science.

\appendix

\section{ SOLVING THE TWO-CENTER-OSCILLATOR EIGENVALUE PROBLEM }
\label{tco}

For a given interwell separation $d$, the single-particle levels of $H_{\rm TCO}$ [ Eq.\ (\ref{hsp}) ] 
are obtained by numerical diagonalization in a basis consisting of the eigenstates of the auxiliary 
hamiltonian:
\begin{equation}
H_0=\frac{{\bf p}^2}{2m^*} + \frac{1}{2} m^* \omega_y^2 y^2
      + \frac{1}{2} m^* \omega_{xk}^2 x_k^{\prime 2}+h_k~.
\label{h0}
\end{equation}
The eigenvalue problem associated with the auxiliary hamiltonian [Eq.\ (\ref{h0})] 
is separable in the $x$ and $y$ variables, i.e., the wave functions are written as 
\begin{equation}
\varphi_i (x,y)= X_\mu (x) Y_n (y),
\label{spphi}
\end{equation}
with $i \equiv \{\mu,n\}$, $i=1,2,\ldots,K$. $K$ specifies the size of the single-particle
basis.

The $Y_n (y)$ are the eigenfunctions of a one-dimensional oscillator, and the
$X_\mu (x \leq 0)$ or $X_\mu (x>0)$ can be expressed through the parabolic
cylinder functions $U[\gamma_k, (-1)^k \xi_k]$, where
$\xi_k = x^\prime_k \sqrt{2m^* \omega_{xk}/\hbar}$, 
$\gamma_k=(-E_x+h_k)/(\hbar \omega_{xk})$, 
and $E_x=(\mu+0.5)\hbar \omega_{x1} + h_1$ denotes the $x$-eigenvalues.
The matching conditions at $x=0$ for the left and 
right domains yield the $x$-eigenvalues and the eigenfunctions
$X_\mu (x)$. The $n$ indices are integer. The number of $\mu$ indices is 
finite; they are in general real numbers.

\section{THE CI METHOD AS ADAPTED TO THE DOUBLE-DOT CASE}
\label{cime}

As aforementioned, we use the method of configuration interaction for determining 
the solution of the many-body problem specified by the Hamiltonian (\ref{mbhd}). 

In the CI method, one writes the many-body wave function 
$\Phi^{\rm CI}_N ({\bf r}_1, {\bf r}_2, \ldots , {\bf r}_N)$ as a linear
superposition of Slater determinants 
$\Psi^N({\bf r}_1, {\bf r}_2, \ldots , {\bf r}_N)$ that span the many-body
Hilbert space and are constructed out of the single-particle 
{\it spin-orbitals\/} 
\begin{equation}
\chi_j (x,y) = \varphi_j (x,y) \alpha, \mbox{~~~if~~~} 1 \leq j \leq K,
\label{chi1}
\end{equation}
and 
\begin{equation}
\chi_j (x,y) = \varphi_{j-K} (x,y) \beta, \mbox{~~~if~~~} K < j \leq 2K, 
\label{chi2}
\end{equation}
where $\alpha (\beta)$ denote up (down) spins, and the spatial orbitals $\varphi_j(x,y)$ 
are defined in Eq.\ (\ref{spphi}). Namely
\begin{equation}
\Phi^{\rm CI}_{N,q} ({\bf r}_1, \ldots , {\bf r}_N) = 
\sum_I C_I^q \Psi^N_I({\bf r}_1, \ldots , {\bf r}_N),
\label{mbwf}
\end{equation}
where 
\begin{equation}
\Psi^N_I = \frac{1}{\sqrt{N!}}
\left\vert
\begin{array}{ccc}
\chi_{j_1}({\bf r}_1) & \dots & \chi_{j_N}({\bf r}_1) \\
\vdots & \ddots & \vdots \\
\chi_{j_1}({\bf r}_N) & \dots & \chi_{j_N}({\bf r}_N) \\
\end{array}
\right\vert,
\label{detexd}
\end{equation}
and the master index $I$ counts \cite{note4} the number of arrangements $\{j_1,j_2,\ldots,j_N\}$ under 
the restriction that $1 \leq j_1 < j_2 <\ldots < j_N \leq 2K$. Of course, $q=1,2,\ldots$ counts
the excitation spectrum, with $q=1$ corresponding to the ground state.

The many-body Schr\"{o}dinger equation 
\begin{equation}
{\cal H} \Phi^{\rm CI}_{N,q} = E^{\rm CI}_{N,q} \Phi^{\rm CI}_{N,q}
\label{mbsch}
\end{equation}
transforms into a matrix diagonalizatiom problem, which yields the coefficients 
$C_I^q$ and the eigenenergies $E^{\rm CI}_{N,q}$. Because the
resulting matrix is sparse, we implement its numerical diagonalization 
employing the well known ARPACK solver \cite{arpack}. Convergence of the many-body 
solutions is guaranteed by using a large enough value for the dimension $K$ of the 
single-particle basis; see Appendix \ref{tco}. The attribute ``full'' is usually 
used for such well converged CI solutions, which naturally contain all possible
$np-nh$ basis Slater determinants.

The matrix elements $\langle \Psi_N^{I} | {\cal H} | \Psi_N^{J} \rangle$
between the basis determinants [see Eq.\ (\ref{detexd})] are calculated using
the Slater rules \cite{yann09,szabobook}. Naturally, an important ingredient in this
respect are the matrix elements of the two-body interaction,
\begin{equation}
\int_{-\infty}^{\infty} \int_{-\infty}^{\infty} d{\bf r}_1 d{\bf r}_2
\varphi^*_i({\bf r}_1) \varphi^*_j({\bf r}_2) V({\bf r}_1,{\bf r}_2)
\varphi_k({\bf r}_1) \varphi_l({\bf r}_2),
\label{clme}
\end{equation}
in the basis formed out of the single-particle spatial orbitals 
$\varphi_i({\bf r})$, $i=1,2,\ldots,K$ [Eq.\ (\ref{spphi})]. In our approach, 
these matrix elements are determined numerically and stored separately.

The Slater determinants $\Psi^N_I$ [see Eq.\ (\ref{detexd})] conserve the
third projection $S_z$,  but not the square $\hat{\bf S}^2$ of the total spin. 
However, because $\hat{\bf S}^2$ commutes with the many-body Hamiltonian, the
CI solutions are automatically eigenstates of $\hat{\bf S}^2$ with eigenvalues
$S(S+1)$. After diagonalization, these eigenvalues are determined by
applying $\hat{\bf S}^2$ onto $\Phi^{\rm CI}_{N,q}$ and using the relation \cite{paunczbook}
\begin{equation}
\hat{{\bf S}}^2 \Psi^N_I = 
\left [(N_\alpha - N_\beta)^2/4 + N/2 + \sum_{i<j} \varpi_{ij} \right ] 
\Psi^N_I,
\end{equation}
where the operator $\varpi_{ij}$ interchanges the spins of fermions $i$ and 
$j$ provided that these spins are different; $N_\alpha$ and $N_\beta$ denote 
the number of spin-up and spin-down fermions, respectively.

\section{SINGLE-PARTICLE DENSITIES AND CONDITIONAL PROBABILITY DISTRIBUTIONS FROM CI
WAVE FUNCTIONS}
\label{spdcpd}

The single-particle density (charge density) is the expectation value of the one-body operator
\begin{equation}
\rho({\bf r}) = \langle \Phi^{\rm CI}_N 
\vert  \sum_{i=1}^N \delta({\bf r}-{\bf r}_i)
\vert \Phi^{\rm CI}_N \rangle,
\label{elden}
\end{equation}
where, as previously, $\Phi^{\rm CI}_N$ denotes the many-body (multi-determinantal) CI wave function
(henceforth, we will drop the $q$ subscript).
For the spin-resolved densities, the expression above is modified as follows:
\begin{equation}
\rho_{\sigma}({\bf r}) = \langle \Phi^{\rm CI}_N
\vert  \sum_{i=1}^N \delta({\bf r}-{\bf r}_i) \delta_{\sigma \sigma_i}
\vert \Phi^{\rm CI}_N \rangle,
\label{eldensig}
\end{equation}
where $\sigma$ denotes either an up or a down spin.

Naturally several distinct spin structures can correspond to the same charge density. The spin 
structure associated with a specific CI wave function can be determined uniquely with the help of the 
spin-resolved densities in conjunction with the many-body spin-resolved CPDs.

The spin-resolved CPDs (referred to also as spin-resolved two-point anisotropic correlation functions) 
yield the conditional probability distribution of finding another fermion with up (or down) spin 
$\sigma$ at a position ${\bf r}$, assuming that a given fermion with up (or down) spin $\sigma_0$ is fixed 
at ${\bf r_0}$. In detail, a spin-resolved CPD is defined as
\begin{equation}
P_{\sigma\sigma_0}({\bf r}, {\bf r}_0)= 
\langle \Phi^{\rm CI}_N |
\sum_{i \neq j} \delta({\bf r} - {\bf r}_i) \delta({\bf r}_0 - {\bf r}_j)
\delta_{\sigma \sigma_i} \delta_{\sigma_0 \sigma_j}
|\Phi^{\rm CI}_N \rangle.
\label{tpcorr}
\end{equation}


\begin{thebibliography}{99}
\bibitem{kouw07}
R. Hanson, L.P. Kouwenhoven, J.R. Petta, S. Tarucha, and L.M.K. Vandersypen,
Spins in few-electron quantum dots,
Rev. Mod. Phys. {\bf 79}, 1217 (2007).
\url{https://doi.org/10.1103/RevModPhys.79.1217}
\bibitem{dzur13}
F.A. Zwanenburg, A.S. Dzurak, A. Morello, M.Y. Simmons, L.C.L. Hollenberg, G. Klimeck, 
S. Rogge, S.N. Coppersmith, and M.A. Eriksson,
Silicon quantum electronics,
Rev. Mod. Phys. {\bf 85}, 961 (2013).
\url{https://doi.org/10.1103/RevModPhys.85.961}
\bibitem{marc10}
C. Barthel, M. Kj{\ae}rg{\aa}rd, J. Medford, M. Stopa, C.M. Marcus, M.P. Hanson, and A.C. Gossard,
Fast sensing of double-dot charge arrangement and spin state with a radio-frequency sensor quantum dot,
Phys. Rev. B {\bf 81}, 161308(R) (2010).
\url{https://doi.org/10.1103/PhysRevB.81.161308}
\bibitem{dzur10}
Single-shot readout of an electron spin in silicon
A. Morello, J.J. Pla, F.A. Zwanenburg, K.W. Chan, K.Y. Tan, H. Huebl, M. M\"{o}tt\"{o}nen, 
C.D. Nugroho, C. Yang, J.A. van Donkelaar, A.D.C. Alves, D.N. Jamieson, C. C. Escott, L.C.L. Hollenberg, 
R.G. Clark, and A.S. Dzurak, 
Nature {\bf 467}, 687 (2010).
\url{https://doi.org/10.1038/nature0939}
\bibitem{yaco18}
L.A. Orona, J.M. Nichol, S.P. Harvey, C.G.L. B{\o}ttcher, S. Fallahi, G.C. Gardner, 
M.J. Manfra, and A. Yacoby,
Readout of singlet-triplet qubits at large magnetic field gradients,
Phys. Rev. B {\bf 98}, 125404 (2018).
DOI: \url{https://doi.org/10.1103/PhysRevB.98.125404}
\bibitem{marc13}
J. Medford, J. Beil, J.M. Taylor, S.D. Bartlett, A.C. Doherty, E.I. Rashba, 
D.P. DiVincenzo, H. Lu, A.C. Gossard, and C.M. Marcus,
Self-consistent measurement and state tomography of an exchange-only spin qubit,
Nature Nanotech. {\bf 8}, 654 (2013).
DOI: \url{ https://doi.org/10.1038/nnano.2013.168}
\bibitem{pett21}
A.R. Mills, C.R. Guinn, M.J. Gullans, A.J. Sigillito, M.M. Feldman, E. Nielsen, and J.R. Petta,
Two-qubit silicon quantum processor with operation fidelity exceeding 99\%,
Sci. Adv. {\bf 8}, eabn5130 (2022).
DOI: \url{https://doi.org/10.1126/sciadv.abn5130}
\bibitem{copp12}
Z. Shi, C.B. Simmons, J.R. Prance, J.K. Gamble, T.S. Koh, Y.-P. Shim, X. Hu, 
D.E. Savage, M.G. Lagally, M.A. Eriksson, M. Friesen, and S.N. Coppersmith,
Fast Hybrid Silicon Double-Quantum-Dot Qubit,
Phys. Rev. Lett. {\bf 108}, 140503 (2012).
DOI: \url{https://doi.org/10.1103/PhysRevLett.108.140503}
\bibitem{copp14.2}
Z. Shi, C.B. Simmons, D.R. Ward, J.R. Prance, X. Wu, T.S.  Koh, J. K. Gamble, 
D.E. Savage, M.G. Lagally, M. Friesen, S.N. Coppersmith, and M.A. Eriksson,
Fast coherent manipulation of three-electron states in a double quantum dot,
Nat Commun 5, 3020 (2014). 
DOI: \url{https://doi.org/10.1038/ncomms4020}
\bibitem{copp15}
D. Kim, D.R. Ward, C.B Simmons, D.E. Savage, M.G. Lagally, M. Friesen, 
S.N. Coppersmith, and M.A. Eriksson,
High-fidelity resonant gating of a silicon-based quantum dot hybrid qubit,
npj Quantum Inf. {\bf 1}, 15004 (2015). 
DOI: \url{https://doi.org/10.1038/npjqi.2015.4}
\bibitem{cao16}
G. Cao, H.-O. Li, G.-D. Yu, B.-C. Wang, B.-B. Chen, X.-X. Song, M. Xiao, G.-C. Guo, 
H.-W. Jiang, X. Hu, and G.-P. Guo,
Tunable Hybrid Qubit in a GaAs Double Quantum Dot,
Phys. Rev. Lett. {\bf 116}, 086801 (2016).
DOI: \url{https://doi.org/10.1103/PhysRevLett.116.086801}
\bibitem{copp17}
B. Thorgrimsson, D. Kim, Y.-C. Yang, L.W. Smith, C.B. Simmons, D.R. Ward, R.H. Foote, 
J. Corrigan, D.E. Savage, M.G. Lagally, M. Friesen, S.N. Coppersmith, and M.A. Eriksson,
Extending the coherence of a quantum dot hybrid qubit,
npj Quantum Inf. {\bf 3}, 32 (2017). 
DOI: \url{https://doi.org/10.1038/s41534-017-0034-2}
\bibitem{cao17}
B.-B. Chen, B.-C. Wang, G. Cao, H.-O. Li, M. Xiao, G.-C. Guo, H.-W. Jiang, 
X. Hu, and G.-P. Guo, 
Spin Blockade and Coherent Dynamics of High-Spin States in a Three-Electron Double Quantum Dot, 
Phys. Rev. B {\bf 95}, 035408 (2017).
DOI: \url{https://doi.org/10.1103/PhysRevB.95.035408}
\bibitem{erca21}
J. Corrigan, J.P. Dodson, H.E. Ercan, J.C. Abadillo-Uriel, B. Thorgrimsson, T.J. Knapp, 
N. Holman, Th. McJunkin, S.F. Neyens, E.R. MacQuarrie, R.H. Foote, L.F. Edge, M. Friesen, 
S.N. Coppersmith, and M.A. Eriksson,
Coherent Control and Spectroscopy of a Semiconductor Quantum Dot Wigner Molecule,
Phys. Rev. Lett. {\bf 127}, 127701 (2021).
DOI: \url{https://doi.org/10.1103/PhysRevLett.127.127701}
\bibitem{kim21}
W. Jang, M-K. Cho, H. Jang, J. Kim, J. Park, G. Kim, B. Kang, H. Jung, V. Umansky, and D. Kim,
Single-Shot Readout of a Driven Hybrid Qubit in a GaAs Double Quantum Dot,
Nano Lett. {\bf 21}, 4999 (2021).
DOI: \url{https://doi.org/10.1021/acs.nanolett.1c00783}
\bibitem{erca21.2}
H.E. Ercan, S.N. Coppersmith, and M. Friesen,
Strong electron-electron interactions in Si/SiGe quantum dots,
arXiv:2105.10645. \url{https://arxiv.org/abs/2105.10645}
\bibitem{urie21}
J.C. Abadillo-Uriel, B. Martinez, M. Filippone, and Y.-M. Niquet,
Two-body Wigner molecularization in asymmetric quantum dot spin qubits,
Phys. Rev. B {\bf 104}, 195305 (2021).
\url{https://doi.org/10.1103/PhysRevB.104.195305}
\bibitem{yann99}
C. Yannouleas and U. Landman,
Spontaneous Symmetry Breaking in Single and Molecular Quantum Dots,
Phys. Rev. Lett. {\bf 82}, 5325 (1999); Erratum Phys. Rev. Lett. {\bf 85}, 2220 (2000).
DOI: \url{https://doi.org/10.1103/PhysRevLett.82.5325},
DOI: \url{https://doi.org/10.1103/PhysRevLett.85.2220}
\bibitem{grab99}
R. Egger, W. H\"{a}usler, C.H. Mak, and H. Grabert,
Crossover from Fermi Liquid to Wigner Molecule Behavior in Quantum Dots,
Phys. Rev. Lett. {\bf 82}, 3320 (1999); Erratum Phys. Rev. Lett. {\bf 83}, 462 (1999).
\url{https://doi.org/10.1103/PhysRevLett.82.3320}
\bibitem{yann00}
C. Yannouleas and U. Landman,
Formation and control of electron molecules in artificial atoms: 
Impurity and magnetic-field effects,
Phys. Rev. B {\bf 61}, 15895 (2000).
DOI: \url{https://doi.org/10.1103/PhysRevB.61.15895}
\bibitem{yann00.2}
C. Yannouleas and U. Landman,
Collective and Independent-Particle Motion in Two-Electron Artificial Atoms,
Phys. Rev. Lett. {\bf 85}, 1726 (2000).
DOI: \url{https://doi.org/10.1103/PhysRevLett.85.1726}
\bibitem{boni01}
A.V. Filinov, M. Bonitz, and Yu.E. Lozovik,
Wigner Crystallization in Mesoscopic 2D Electron Systems,
Phys. Rev. Lett. {\bf 86}, 3851 (2001).
\url{https://doi.org/10.1103/PhysRevLett.86.3851}
\bibitem{yann02}
C. Yannouleas and U. Landman,
Magnetic-field manipulation of chemical bonding in artificial molecules,
Int. J. Quantum Chem. {\bf 90}, 699 (2002).
DOI: \url{https://doi.org/10.1002/qua.980};
Strongly correlated wavefunctions for artificial atoms and molecules,
J. Phys.: Condens. Matter {\bf 14}, L591 (2002).
DOI: \url{https://doi.org/10.1088/0953-8984/14/34/101}
\bibitem{mikh02}
S.A. Mikhailov,
Quantum-dot lithium in zero magnetic field: Electronic properties, thermodynamics, 
and Fermi liquid–Wigner solid crossover in the ground state,
Phys. Rev. B {\bf 65}, 115312 (2002).
\url{https://doi.org/10.1103/PhysRevB.65.115312}
\bibitem{szaf03}
M.B. Tavernier, E. Anisimovas, F.M. Peeters, B.Szafran, J. Adamowski, and S. Bednarek,
Four-electron quantum dot in a magnetic field,
Phys. Rev. B {\bf 68}, 205305 (2003).
DOI: \url{https://doi.org/10.1103/PhysRevB.68.205305}
\bibitem{cift05}
O. Ciftja and M.G. Faruk,
Two-dimensional quantum-dot helium in a magnetic field: Variational theory,
Phys. Rev. B {\bf 72}, 205334 (2005).
DOI: \url{https://doi.org/10.1103/PhysRevB.72.205334}
\bibitem{yann06}
C. Ellenberger, T. Ihn, C. Yannouleas, U. Landman, K. Ensslin, D. Driscoll, and A.C. Gossard,
Excitation Spectrum of Two Correlated Electrons in a Lateral Quantum Dot with 
Negligible Zeeman Splitting,
Phys. Rev. Lett. {\bf 96}, 126806 (2006).
DOI: \url{https://doi.org/10.1103/PhysRevLett.96.126806}
\bibitem{yann06.2}
T. Ihn, C. Ellenberger, K. Ensslin, C. Yannouleas, U. Landman, D.C. Driscoll, and A.C. Gossard,
Quantum dots based on parabolic quantum wells: Importance of electronic correlations,
Int. J. Mod. Phys. {\bf 21}, 1316 (2007).
DOI: \url{https://doi.org/10.1142/S0217979207042781}
\bibitem{yann07}
C. Yannouleas and U. Landman,
Symmetry breaking and quantum correlations in finite systems:
studies of quantum dots and ultracold Bose gases and related nuclear and chemical methods,
Rep. Prog. Phys. {\bf 70}, 2067 (2007).
DOI: \url{https://doi.org/10.1088/0034-4885/70/12/R02}
\bibitem{yann08}
L.O. Baksmaty, C. Yannouleas, and U. Landman,
Nonuniversal Transmission Phase Lapses through a Quantum Dot:
An Exact Diagonalization of the Many-Body Transport Problem,
Phys. Rev. Lett. {\bf 101}, 136803 (2008).
DOI: \url{https://doi.org/10.1103/PhysRevLett.101.136803}
\bibitem{ront08}
S. Kalliakos, M. Rontani, V. Pellegrini, C.P. Garc\'{i}a, A. Pinczuk, 
G. Goldoni, E. Molinari, L.N. Pfeiffer, and K.W. West, 
A molecular state of correlated electrons in a quantum dot,
Nature Physics {\bf 4}, 467 (2008).
\url{https://doi.org/10.1038/nphys944}
\bibitem{wign34}
E. Wigner,
On the Interaction of Electrons in Metals,
Phys. Rev. {\bf 46}, 1002 (1934).
\url{https://doi.org/10.1103/PhysRev.46.1002}
\bibitem{copp14}
D. Kim, Z. Shi, C.B. Simmons, D.R. Ward, J.R. Prance, T.S. Koh, J.K. Gamble, D.E. Savage, 
M.G. Lagally, M. Friesen, S.N. Coppersmith, and M.A. Eriksson,
Quantum control and process tomography of a semiconductor quantum dot hybrid qubit,
Nature {\bf 511}, 70 (2014).
DOI: \url{https://doi.org/10.1038/nature13407}
\bibitem{vinc15}
M.A. Bakker, S. Mehl, T. Hiltunen, A. Harju, and D.P. DiVincenzo,
Validity of the single-particle description and charge noise resilience for multielectron
quantum dots,
Phys. Rev. B {\bf 91}, 155425 (2015).
DOI: \url{https://doi.org/10.1103/PhysRevB.91.155425}
\bibitem{marc96}
See an early review on quantum dots: 
L.P. Kouwenhoven, C.M. Marcus, P.L. McEuen, S. Tarucha, R.M. Westervelt. and N.S. Wingreen,
Electron transport in quantum dots, in {\it Mesoscopic Electron Transport\/}, 
edited by L.L. Sohn, L.P. Kouwenhoven, and G. Sch.\"{o}n,
NATO ASI Series (Series E: Applied Sciences), vol 345
(Springer, Dordrecht, 1997) pp. 105–214.
DOI: \url{https://doi.org/10.1007/978-94-015-8839-3_4}
\bibitem{staf94}
C.A. Stafford and S. Das Sarma,
Collective Coulomb blockade in an array of quantum dots: A Mott-Hubbard approach,
Phys. Rev. Lett. {\bf 72}, 3590 (1994).
DOI: \url{https://doi.org/10.1103/PhysRevLett.72.3590}
\bibitem{loss99}
Coupled quantum dots as quantum gates
G. Burkard, D. Loss, and D.P. DiVincenzo,
Phys. Rev. B {\bf 59}, 2070 (1999).
DOI: \url{https://doi.org/10.1103/PhysRevB.59.2070}
\bibitem{ferr14}
E. Ferraro, M. De Michielis, G. Mazzeo, M. Fanciulli, and E. Prati,
Effective Hamiltonian for the hybrid double quantum dot qubit,
Quantum Inf. Process. {\bf 13} 1155 (2014).
DOI: \url{10.1007/s11128-013-0718-2}
\bibitem{dass11}
S. Yang, X. Wang, and S. Das Sarma,
Generic Hubbard model description of semiconductor quantum-dot spin qubits,
Phys. Rev. B {\bf 83}, 161301(R) (2011).
DOI: \url{https://doi.org/10.1103/PhysRevB.83.161301}
\bibitem{dass11.2}
S. Das Sarma, X. Wang, and S. Yang,
Hubbard model description of silicon spin qubits: Charge stability diagram and 
tunnel coupling in Si double quantum dots,
Phys. Rev. B {\bf 83}, 235314 (2011).
DOI: \url{https://doi.org/10.1103/PhysRevB.83.235314}
\bibitem{burk17}
M. Russ and G. Burkard,
Three-electron spin qubits,
J. Phys.: Condens. Matter {\bf 29}, 393001 (2017).  
DOI: \url{https://doi.org/10.1088/1361-648X/aa761f}
\textcolor{black}{
\bibitem{note3}
Each site in these {\it two-site} models \cite{copp12,staf94,loss99,dass11,dass11.2,ferr14,burk17}
is associated with a given dot in the double-dot device. It is then unavoidable to use the single-particle 
orbital eigenstates of the separated dots (see Fig.\ 3 in the Supplemental Material of Ref.\ \cite{copp12}) 
in order to construct an approximate many-body wave function for the double-dot qubit.
This approach cannot describe the many-body wave function in the case of strong
$e-e$ correlations and Wigner-molecule formation, when (due to the abscence of rotational symmetry
in the double dot) the electrons localize within each dot at fixed positions that can only be extracted 
by plotting the spin-unresolved and spin-resolved densities (see Sec.\ \ref{sec:char1} and Sec.\ 
\ref{sec:spnstr}) as a derivative result from a full CI calculation. In the case of strong WM formation, 
the localized electrons (or fermionic particles in general) can be approximately described by 
{\it displaced Gaussians\/}, and a mapping can be established between CI densities (as well as
conditional probabilities; see Appendix \ref{spdcpd} for definitions) and 
spin eigenfunctions \big(see Ref.\ \cite{paunczbook} and Eqs.\ (\ref{spin1}) and (\ref{spin2})\big).
Such a mapping was discussed in detail in Ref.\ \cite{yann09} for four electrons in a double quantum dot 
and for three ultracold fermionic atoms in a double optical trap in C. Yannouleas, B.B. Brandt, and 
U. Landman, Ultracold few fermionic atoms in needle-shaped double wells: 
spin chains and resonating spin clusters from microscopic Hamiltonians emulated via antiferromagnetic 
Heisenberg and $t-J$ models, New J. Phys. {\bf 18}, 073018 (2016),
DOI: \url{https://doi.org/10.1088/1367-2630/18/7/073018}.
This strong-WM $\leftrightarrow$ spin-eigenfunctions mapping implies that the strong-WM physics can be 
interpreted qualitatively using the limiting case (for large $U/t$) of the Hubbard model known as the 
Heisenberg model. At a minimum, however, the number of sites in such a Heisenberg modeling is $N$ (the 
number of localized fermions), and even larger in the case of a resonance between the left and right 
potential wells; for details, see again the two references mentioned above in this footnote.
Elaborating further on such a Heisenberg-model interpretation in the context of a double-dot qubit
is beyond the scope of the present paper.
}
\bibitem{yann07.2}
Yuesong Li, C. Yannouleas, and U. Landman,
Three-electron anisotropic quantum dots in variable magnetic fields: 
Exact results for excitation spectra, spin structures, and entanglement,
Phys. Rev. B {\bf 76}, 245310 (2007); Erratum Phys. Rev. B {\bf 81}, 049902 (2010).
DOI: \url{https://doi.org/10.1103/PhysRevB.76.245310},
DOI: \url{https://doi.org/10.1103/PhysRevB.81.049902}
\bibitem{yann09}
Ying Li, C. Yannouleas, and U. Landman,
Artificial quantum-dot helium molecules: Electronic spectra, spin structures, 
and Heisenberg clusters,
Phys. Rev. B {\bf 80}, 045326 (2009).
DOI: \url{https://doi.org/10.1103/PhysRevB.80.045326}
\bibitem{shav98}
For a historical perspective of the CI method in the field of quantum chemistry, see
I. Shavitt, 
The history and evolution of configuration interaction,
Molec. Phys. {\bf 94}, 3 (1998).
DOI: \url{https://doi.org/10.1080/002689798168303}
\bibitem{note1}
For a detailed discussion of these two methodologies in the context of
QDs, see the review article in Ref.\ \cite{yann07}.
\bibitem{kouw01}
See, e.g., L.P. Kouwenhoven, D.G. Austing, and S. Tarucha,
Few-electron quantum dots,
Rep. Prog. Phys. {\bf 64}, 701 (2001).
DOI: \url{https://doi.org/10.1088/0034-4885/64/6/201}
\bibitem{paunczbook}
R. Pauncz,
{\it The Construction of Spin Eigenfunctions: An Exercise Book\/}
(New York, Kluwer Academic/Plenum Publishers, 2000).
\bibitem{note2}
See Sec. IV and Fig.\ 2 in Ref.\ \cite{yann09}.
\bibitem{vinc00}
D.P. DiVincenzo, D. Bacon, J. Kempe, G. Burkard, and K.B. Whaley, 
Universal quantum computation with the exchange interaction,
Nature {\bf 408}, 339 (2000).
\url{https://doi.org/10.1038/35042541}
\bibitem{note5}
The energy spectra do not depend on the total-spin projection $S_z$. However, the many-body wave 
functions and the associated spin structures do depend on $S_z$.
\bibitem{suzubook}
See Y. Suzuki and K. Varga,
{\it Stochastic Variational Approach to Quantum-Mechanical Few-Body Problems\/}
(Springer, Berlin, 1998); see also Ying Li, Ph.D. dissertation, Georgia
Institute of Technology (2009), Ch. 3.4.2. 
\url{http://hdl.handle.net/1853/28186}.
\bibitem{cao13}
G. Cao, H.O. Li, T. Tu, L. Wang, C. Zhou, M. Xiao, G.C. Guo,
H.W. Jiang, and G.-P. Guo,
Ultrafast universal quantum control of a quantum-dot charge qubit using
Landau-Zener-St\"{u}ckelberg interference,
Nat. Commun. {\bf 4}, 1401 (2013).
\url{https://doi.org/10.1038/ncomms2412}
\bibitem{copp12.2}
T.S. Koh, J.K. Gamble, M. Friesen, M.A. Eriksson, and S.N. Coppersmith,
Pulse-Gated Quantum-Dot Hybrid Qubit
Phys. Rev. Lett. {\bf 109}, 250503 (2012).
\url{https://doi.org/10.1103/PhysRevLett.109.250503}  
\bibitem{burk13}
H. Ribeiro, J.R. Petta, and G. Burkard,
Interplay of charge and spin coherence in Landau-Zener-St\"{u}ckelberg-Majorana interferometry,
Phys. Rev. B {\bf 87}, 235318 (2013).
\url{https://doi.org/10.1103/PhysRevB.87.235318}
\bibitem{cao17.2} 
B.-C. Wang, G. Cao, H.-O. Li, M. Xiao, G.-C. Guo, X. Hu, H.-W. Jiang, and G.-P. Guo,
Tunable Hybrid Qubit in a Triple Quantum Dot,
Phys. Rev. Applied {\bf 8}, 064035 (2017)
\url{https://doi.org/10.1103/PhysRevApplied.8.064035}
\bibitem{yann22}  
C. Yannouleas and U. Landman,
unpublished.
\bibitem{note6}   
See Sec.\ S5 in the Supplemental Material of Ref.\ \cite{erca21}.
\bibitem{veld20}  
N.W. Hendrickx, D.P. Franke, A. Sammak, G. Scappucci, and M. Veldhorst, 
Fast two-qubit logic with holes in germanium,
Nature {\bf 577}, 487 (2020).
\url{https://doi.org/10.1038/s41586-019-1919-3}
\bibitem{veld21}
N.W. Hendrickx, W.I.L. Lawrie, M. Russ, F. van Riggelen, S.L. de Snoo, R.N. Schouten, 
A. Sammak, G. Scappucci, and Menno Veldhorst,
A four-qubit germanium quantum processor. 
Nature {\bf 591}, 580 (2021). 
\url{https://doi.org/10.1038/s41586-021-03332-6}
\textcolor{black}{
\bibitem{note4}
The question of the optimal set of expansion coefficients $C_I^q$ (in the sense of providing
the {\it most rapid numerical convergence}) in Eq.\ (\ref{mbwf})
was answered via the introduction of the concept of natural orbitals in 
P.-O. L\"{o}wdin, 
Quantum Theory of Many-Particle Systems. I. Physical Interpretations by Means of Density Matrices, 
Natural Spin-Orbitals, and Convergence Problems in the Method of Configurational Interaction,
Phys. Rev. {\bf 97}, 1474 (1955),
DOI: \url{https://doi.org/10.1103/PhysRev.97.1474}
and in
P.-O. L\"{o}wdin and H. Shull,
Natural Orbitals in the Quantum Theory of Two-Electron Systems,
Phys. Rev. {\bf 101}, 1730 (1956),
DOI: \url{https://doi.org/10.1103/PhysRev.101.1730}; 
see also p. 8 in Ref.\ \cite{shav98} and 
E.R. Davidson,
Properties and Uses of Natural Orbitals,
Rev. Mod. Phys. {\bf 44}, 451 (1972),
DOI: \url{https://doi.org/10.1103/RevModPhys.44.451}.
For a use of this optimal expansion in the case of two fermions confined in a double-well potential, 
see B.B. Brandt, C. Yannouleas, and U. Landman,
Double-Well Ultracold-Fermions Computational Microscopy: Wave-Function Anatomy of 
Attractive-Pairing and Wigner-Molecule Entanglement and Natural Orbitals,
Nano Lett. {\bf 15}, 7105 (2015), DOI: \url{https://doi.org/10.1021/acs.nanolett.5b03199};
in particular, see therein the Section ``Quantifying  Entanglement  Using  the  von  Neumann
Entropy  as  a  Measure  and  the  Natural  Orbitals'' and the insets in Fig.\ 4, where the
occupations of the natural orbitals are plotted for several instances. For a similar analysis
for two electrons in an elliptic quantum dot, see  
C. Yannouleas and U. Landman,
Electron and boson clusters in confined geometries: Symmetry breaking in quantum dots and harmonic traps,
Proc. Natl. Acad. Sci. (USA) {\bf 103}, 10600 (2006),
DOI: \url{https://doi.org/10.1073/pnas.0509041103};
Symmetry breaking and Wigner molecules in few-electron quantum dots,
phys. stat. sol. (a) {\bf 203}, 1160 (2006),
DOI: \url{https://doi.org/10.1002/pssa.200566197}.
}
\bibitem{arpack}
R. B. Lehoucq, D. C. Sorensen, and C. Yang, 
{\it ARPACK USERS' GUIDE: Solution of large-scale eigenvalue 
problems with implicitly restarted ARNOLDI methods\/}
(SIAM, Philadelphia, 1998).
\bibitem{szabobook} 
A. Szabo and N.S. Ostlund,
{\it Modern Quantum Chemistry\/}
(New York, McGraw-Hill, 1989) Chap. 4.
\end{thebibliography}
\end{document}